\documentclass[%
 reprint, nofootinbib,
 amsmath,amssymb,
 aps, prd,
]{revtex4-2}

\usepackage[dvipsnames]{xcolor}
\usepackage{appendix}
\usepackage[utf8]{inputenc}
\usepackage{soul}
\usepackage{graphicx}
\usepackage{xcolor}
\usepackage{mathrsfs}
\usepackage{hyperref}
\usepackage{braket}
\usepackage{subcaption}

\usepackage{subcaption}
\usepackage{amsthm}
\usepackage{float}
\usepackage{enumitem}
\usepackage{textcomp}
\usepackage{newunicodechar}
\newunicodechar{ }{\,}
\usepackage{notoccite} 
\hypersetup{
    colorlinks=true,
    linkcolor=magenta,
    citecolor = blue,
    filecolor=magenta,      
    urlcolor=blue,
    pdftitle={Overleaf Example}
    }
\usepackage{media9}
\usepackage{float}
\usepackage{dirtytalk}
\usepackage{mathrsfs}
\usepackage{amsmath,amssymb}
\usepackage{physics}
\usepackage{tikz}
\usetikzlibrary{quantikz2}
\usepackage[export]{adjustbox}
\usepackage{lipsum} 
\graphicspath{{Figures/}}
\usepackage{caption}
\usepackage{mdframed}
\usepackage[font=normal,labelfont=bf]{caption}
\usepackage{subcaption}
\usepackage{mathtools}
\usepackage{bm}
\begin{document}

\preprint{APS/123-QED}

\title{Inflation from Covariant Signature Change: A Geometric Mechanism}

\author{Raghvendra Singh}
 \email{raghvendra@ariel.ac.il}
\author{Sergey Bondarenko}%
 \email{sergeyb@ariel.ac.il}
\affiliation{
 Physics Department, Ariel University, Israel}

\date{\today}

\noindent
\begin{abstract}
  We present a covariant mechanism in which a smooth change of metric signature, from a Euclidean to a Lorentzian regime, drives a finite interval of accelerated expansion. The transition, encoded by a scalar interpolator along a timelike congruence, occurs on a codimension-one hypersurface where the continued metric is degenerate but curvature invariants remain finite, so the surface is curvature-regular. Using this covariant continuation, we rewrite the Einstein tensor of the continued metric as a localized, interpolator-dependent effective source for the post-transition Lorentzian branch, yielding a purely geometric stress tensor supported near the crossing. In the Lorentzian regime we derive a model-independent, local criterion for acceleration: inflation persists while the interpolator's slope exceeds a critical value fixed by the extrinsic curvature and the spatial Ricci curvature on the initial hypersurface, and ends when this inequality is first saturated. Standard smooth profiles (tanh, generalized logistic, and power-law/arctan) admit closed-form expressions for the proper-time duration of the accelerated epoch, showing that, for fixed geometric data, the profile shape controls this duration. The construction provides a non-singular, inflaton-free route from a regular Euclidean origin to an early Lorentzian phase of accelerated expansion, in a manner compatible with no--boundary--type boundary conditions.  
\end{abstract}

%
%


\maketitle

\section{Introduction}

The hypothesis that the early universe underwent a transition from a Riemannian (Euclidean) to a Lorentzian metric---commonly called \emph{signature change}---has become a central theme in modern cosmology. By replacing the classical big-bang singularity with a smooth Euclidean phase, signature change offers a route to singularity avoidance, provides a natural seed for an inflationary epoch, and clarifies the quantum-to-classical transition of primordial perturbations~\cite{HartleHawking:1983,GibbonsHartle:1990,BojowaldPaily:2012,Zhang:2019rrc,AshtekarLewandowski:2004,Cailleteau:2012}. The idea is rooted in the Hartle--Hawking no-boundary proposal, in which the universe is born as a compact Euclidean manifold that smoothly evolves into a Lorentzian spacetime~\cite{HartleHawking:1983,GibbonsHartle:1990}.

A practical implementation often invokes analytic continuation (Wick rotation). While this prescription is unambiguous in static geometries, naïve extensions to dynamical spacetimes can be ambiguous. Several covariant prescriptions have therefore been developed~\cite{Visser:Wick,Samuel,Dawood,Dawood2,My}. In classical cosmology, the models of Gibbons and Hartle already demonstrate that a smooth Euclidean--Lorentzian transition is feasible~\cite{GibbonsHartle:1990}. Existing approaches differ mainly in how the matching is implemented: continuous or regular models use a metric, or an induced metric in brane-world settings, that becomes degenerate on the transition hypersurface to ensure smoothness~\cite{HartleHawking:1983,BojowaldPaily:2012,Mars:2000gu,Mars:2007zt,Mars:2007wy,Mars:2007fz}, whereas discontinuous models accommodate abrupt changes by imposing junction conditions and allowing for stress--energy layers at the interface~\cite{BojowaldPaily:2012,Hayward}.

Motivated by the no-boundary path-integral picture, one may regard a compact Euclidean cap as providing regular boundary data for the emergent Lorentzian universe. In this work we adopt a covariant, model-independent description of the signature change itself, without committing to a specific microscopic completion~\cite{Dawood,Dawood2,My}. The transition is encoded by a smooth interpolator $\Theta$ along a preferred timelike congruence $u^{a}$ and occurs on a codimension-one hypersurface where the continued metric becomes degenerate in a controlled way to permit the change of signature. This degeneracy is geometric rather than pathological: for $C^{2}$ profiles with bounded extrinsic curvature and acceleration, curvature invariants (such as the Kretschmann scalar) remain finite at the crossing, so the transition surface is curvature-regular and provides well-posed matching data for the Lorentzian evolution.

Our starting point is the interpolating metric
\begin{equation}
\widehat{g}^{ab} = g^{ab} - \Theta\,u^{a}u^{b},
\end{equation}
where $g^{ab}$ is the Lorentzian metric and $u^{a}$ is a non-vanishing timelike vector field. We interpret the Einstein tensor of the continued metric $\widehat g_{ab}$ as an effective source for $g_{ab}$, defining a localized, purely geometric stress tensor $T_{ab} \equiv \widehat G_{ab} - G_{ab}$ supported near the transition hypersurface. In a nearly homogeneous setting with small shear and heat flux, this tensor admits a fluid decomposition with energy density $\rho$, pressure $P$, and small anisotropic corrections determined entirely by the geometry of the transition surface and by the interpolator $\Theta$.

The central result of this paper is a local, model-independent criterion for accelerated expansion in the Lorentzian regime. Immediately after the transition, the effective equation of state can be written in the form
\begin{equation}
w(t) \equiv \frac{P}{\rho} = -\,\frac{S(t)}{S_c},
\qquad
S(t) \equiv \dot\Theta(t),
\end{equation}
where the \emph{geometric scale} $S_c$ is a positive combination of the expansion scalar and the intrinsic spatial curvature on the transition hypersurface. Inflation occurs whenever the interpolator slope satisfies $S(t) > S_{\rm th}$, with $S_{\rm th} \equiv S_c/3$ corresponding to $w=-1/3$, and ends when this inequality is first saturated. In this sense both the onset and the exit of inflation are controlled purely by local geometric data and by the shape of the interpolator near the signature-change surface.

The paper is organized as follows. Section~\ref{mathematical framework} introduces the covariant formalism for Euclidean--Lorentzian transitions and establishes the geometric foundations of the continuation. Section~\ref{EM tensor} constructs the effective energy--momentum tensor associated with signature change and derives its fluid decomposition, emphasizing the emergence of an intrinsic, purely geometric source. Section~\ref{Inflation} analyzes in detail how accelerated expansion arises from this source, derives the local geometric criterion for inflation, and shows how different smooth interpolators (tanh, generalized logistic, and power-law/arctan) lead to different inflationary durations at fixed geometric threshold. Section~\ref{Conclusion} summarizes our results and discusses their implications for no--boundary--type cosmologies and for future work on perturbations and observational signatures.

\section{Mathematical Framework}\label{mathematical framework}

A covariant mechanism for implementing a change of signature was introduced in
Refs.~\cite{Dawood,Dawood2,My,Bondarenko:2021xvz}.  
The key idea is to promote a background Lorentzian metric $g_{ab}$ to
\begin{equation}\label{eq1}
\widehat{g}_{ab}=g_{ab}+F\,t_{a}t_{b},
\qquad
\widehat{g}^{ab}=g^{ab}-\Theta\,u^{a}u^{b},
\end{equation}
where the scalar function $\Theta$ governs the signature and is related to
$F$ by
\begin{equation}\label{eq2}
F=\frac{\Theta}{1+\Theta}.
\end{equation}
Here $u^{a}$ is the four-velocity of a preferred timelike congruence, 
$t_{a}=g_{ab}\,u^{b}$, and the four-acceleration is
$a^{m}=u^{\ell}\nabla_{\!\ell}u^{m}$. An overdot denotes the derivative
along $u^{a}$, i.e.\ $\dot{X}\equiv u^{m}\nabla_{m}X$.  One then has
\begin{equation}\label{eq3}
\dot{F}=\frac{\dot{\Theta}}{(1+\Theta)^{2}},
\qquad
\nabla_{a}F=-\dot{F}\,t_{a}.
\end{equation}

\begin{figure}[H]
    \centering
    \includegraphics[height=6.5cm,width=0.475\textwidth]{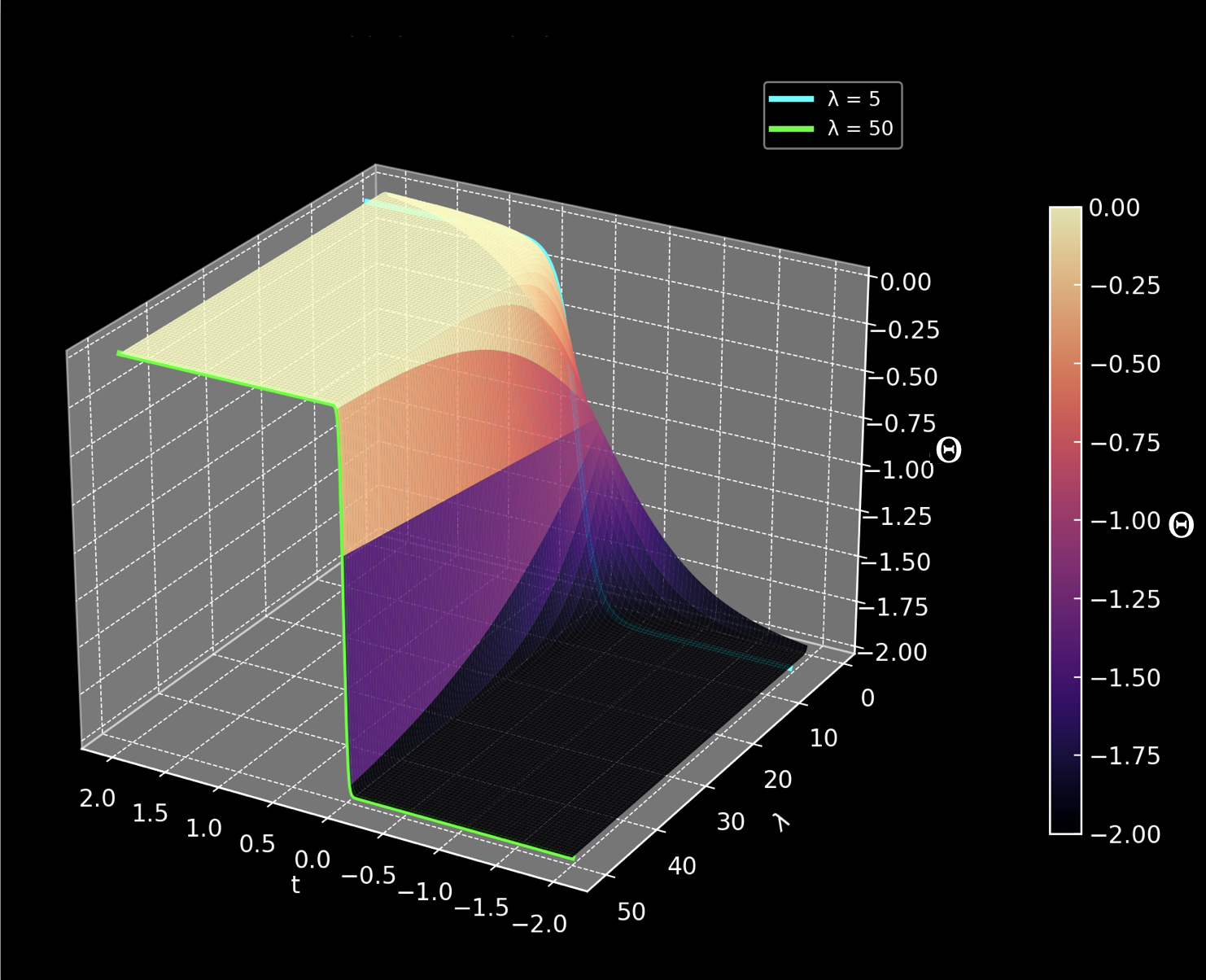}
    \caption{Plot of the signature-changing function 
    $\Theta(\lambda t) = \tanh(\lambda t) - 1$.
    The shaded dark region denotes the Euclidean asymptotic regime ($t\to -\infty$), 
    while the shaded light region denotes the Lorentzian asymptotic regime ($t\to +\infty$).
    For this profile one has $S(t)\equiv\dot\Theta(t)=\lambda\,\mathrm{sech}^2(\lambda t)$,
    so $S(0^+)=\lambda$. Geometric inflation occurs in the non-phantom window 
    $S_{\rm th}<S(t)\le S_c$ (see Sec.~\ref{Inflation}); in particular, 
    $S_{\rm th}<\lambda\le S_c$ corresponds to an inflationary interval just after the transition,
    while larger $\lambda$ produce increasingly sharp interpolations, approaching a step function
    as $\lambda\to\infty$.}
    \label{fig: Signature change}
\end{figure}

\paragraph*{Signature-change hypersurface.}
The equation $\Theta=-1$ defines a hypersurface
$\Sigma_{0}$ across which $F$ diverges and $\widehat{g}^{ab}$ changes
signature: $\Theta>-1$ corresponds to a Lorentzian region, while
$\Theta<-1$ yields a Euclidean region.  A convenient illustrative profile is 
\begin{equation}\label{eq4}
\Theta(\lambda t)=\tanh(\lambda t)-1,
\end{equation}
with $t$ a coordinate adapted to $u^{a}$ and $\lambda>0$ setting the
rapidity (slope) of the transition.  At $t=0$ one has $\Theta=-1$, marking
$\Sigma_{0}$.  As $t\to+\infty$, $\Theta\to0$ and $F\to0$, as $t\to-\infty$, $\Theta\to-2$ and
$F\to2$, giving a distinct Euclidean asymptote (see Fig.~\ref{fig: Signature change}).

\paragraph*{Pure-gravity postulate.}
Following earlier work on signature change and the no-boundary proposal
\cite{HartleHawking:1983,Embacher:1995,HalliwellHertog:2019}, we impose the
vacuum Einstein equations for the \emph{continued} metric:
\begin{equation}\label{eq5}
G_{ab}\!\bigl(\widehat{g}\bigr)=0.
\end{equation}
Thus both the Euclidean ``seed'' phase and the late-time Lorentzian
spacetime are exact vacuum solutions of the full geometry $\widehat g_{ab}$
once the transition layer is left behind.

\paragraph*{Effective-source interpretation.}
An equivalent viewpoint is to treat $g_{ab}$ as the fundamental Lorentzian
metric and interpret the additional term $F\,t_{a}t_{b}$ as inducing an
effective stress–energy tensor for $g_{ab}$.  To make this precise,
write the Einstein tensor of the continued metric as
$\widehat G_{ab}=\widehat G_{ab}(g,\Theta)$ and define
\begin{equation}\label{eq6}
T_{ab}(g,\Theta)\;\equiv\;\widehat G_{ab}(g,\Theta)-G_{ab}(g).
\end{equation}
With this convention the vacuum equation \eqref{eq5} becomes
\begin{equation}\label{Intr8}
G_{ab}(g) + T_{ab}(g,\Theta) = 0,
\end{equation}
so the geometric source $T_{ab}$ enters with the opposite sign to the
usual matter stress tensor.  The tensor $T_{ab}(g,\Theta)$ is localized
near $\Theta=-1$: for smooth profiles such as \eqref{eq4}, $\Theta(t)$ and
its derivatives rapidly approach constants for $|t|\gg\lambda^{-1}$, and
one recovers $T_{ab}\to 0$ and the vacuum Einstein equations for $g_{ab}$
in the far Lorentzian regime.

Near the transition, however, the same relation can be written as
\begin{equation}\label{Intr9}
G_{ab}(g)\simeq -\,T_{ab}\bigl(h_{0}(t),\Theta\bigr),
\end{equation}
where the minus sign is not a matter-source sign, but the geometric stress-balance sign associated with moving the signature-change contribution from the gravitational side to the effective-source side, in the same convention as \cite{My}. Here $h_{ij}(t=0)$ specifies the intrinsic geometry of the
$\Theta=-1$ hypersurface,
\begin{equation}\label{Intr10}
g_{ab}(t=0)=-\,h_{ij}(0)\,\delta^{i}_{a}\delta^{j}_{b}.
\end{equation}
The purely geometrical source $T_{ab}(h_{0},\Theta)$ drives a
finite-duration rearrangement of $g_{ab}$ across $t=0$, producing a short
episode of accelerated expansion before the geometry relaxes back to
vacuum Einstein evolution. From a quantum-cosmological perspective this
source may be interpreted as encoding a change in vacuum state, while the
overall construction realises the no-boundary condition by joining a
compact Euclidean region smoothly to the Lorentzian universe.

\section{Energy-Momentum Tensor in Signature Transition Formalism}\label{EM tensor}

In this section we explore the effective energy–momentum tensor associated with the covariant
signature-change formalism, which bridges Euclidean and Lorentzian regimes within general relativity.
The covariantly continued metric $\widehat{g}^{ab}$ is constructed as
\begin{equation}
\widehat{g}^{ab} = g^{ab} - \Theta\,u^a u^b,
\end{equation}
where $g^{ab}$ is the Lorentzian inverse metric of the original spacetime $(M,g_{ab})$, and $u^a$ is a
non-vanishing timelike vector field ($g_{ab}u^a u^b=-1$) associated with a preferred congruence in the
Lorentzian regime. The scalar interpolator $\Theta$ governs the transition between Euclidean and
Lorentzian geometries, smoothly evolving from $\Theta=-2$ (Euclidean) to $\Theta=0$ (Lorentzian). This
construction allows a covariant signature change and introduces a degenerate hypersurface
$\Sigma_0$, where $\Theta=-1$, marking the transition surface between the two regimes.

We define the effective energy–momentum tensor by
\begin{equation}
T_{ac} \;\equiv\; \widehat{G}_{ac}-G_{ac},
\end{equation}
where $\widehat{G}_{ab}$ and $G_{ab}$ are the Einstein tensors of $\widehat{g}_{ab}$ and $g_{ab}$,
respectively.\footnote{In our construction $T^{a}{}_{c}$ and $T^{ab}$ are not obtained by raising
indices on $T_{ac}$ with $g_{ab}$, but are defined independently as
$T^{a}{}_{c} \equiv \widehat{G}^{a}{}_{c} - G^{a}{}_{c}$. Since the effective fluid variables are defined
by covariant projections of $T_{ab}$ with respect to the preferred congruence $u^{a}$, e.g.
$\rho = T_{ab} u^{a} u^{b}$ and $P = \tfrac{1}{3} h^{ab} T_{ab}$, it is natural to regard
$T_{ab} = \widehat{G}_{ab} - G_{ab}$ as the effective energy–momentum tensor underlying the fluid
interpretation.}
A straightforward but lengthy computation gives
\begin{widetext}
\begin{align}
T_{ab} &= \Theta \Bigg(
      K K_{ab}
    + \nabla_{\mathbf{u}} K_{ab}
    - \frac{1}{1 + \Theta}\, t_a t_b\,\nabla_m a^m
    - 2 a^m t_{(a} K_{b) m}
    - g_{ab}\,\nabla_{\mathbf{u}} K
    - \frac{1}{2} K_{mn} K^{mn} g_{ab}
    - \frac{1}{2} K^2 g_{ab} \nonumber \\
&\qquad\qquad
    - \frac{F}{2}\left(
          t_a t_b\,\nabla_{\mathbf{u}} K
        + t_a t_b\,K_{mn} K^{mn}
        + t_a t_b\,K^2
      \right)
    - \frac{1}{1 + \Theta}\,R\,t_a t_b
    \Bigg)
    - \frac{\dot{\Theta}}{2}\left(
          K h_{ab}
        - K_{ab}
      \right),
\end{align}
\end{widetext}
where $t_a = g_{ab}u^b$, $K_{ab}$ is the extrinsic curvature of the $u^a$-orthogonal slices,
$K = g^{ab}K_{ab}$ its trace, $a^m = u^\ell \nabla_\ell u^m$ the four-acceleration,
$\nabla_{\mathbf{u}} \equiv u^m \nabla_m$, $R$ the Ricci scalar of $g_{ab}$, and
$F=\Theta/(1+\Theta)$ as in Eq.~\eqref{eq2}.

It is convenient to decompose $T_{ab}$ with respect to $u^a$ in the standard imperfect-fluid form,
\begin{align}
T_{ab} = \rho\,t_a t_b + 2 q_{(a} t_{b)} + P\,h_{ab} + \pi_{ab},
\end{align}
where $h_{ab}=g_{ab}+t_a t_b$ projects orthogonally to $u^a$, $\rho$ is the effective
energy density, $q_a$ the heat flux (orthogonal to $u^a$), $P$ the isotropic pressure, and $\pi_{ab}$ the
anisotropic stress tensor, which is symmetric, traceless and transverse,
\begin{equation}
\pi_{ab} = \pi_{(ab)},\qquad \pi^{a}{}_{a}=0,\qquad \pi_{ab}u^b=0.
\end{equation}
These quantities follow from covariant projections of $T_{ab}$.

\paragraph{Energy density and pressure.}
The energy density $\rho$ and pressure $P$ associated with $T_{ab}$ are
\begin{align}\label{eq:rho}
\rho &= \Theta \Bigg(
      \nabla_{\mathbf{u}} K
    + \frac{1}{2} K_{mn} K^{mn}
    + \frac{1}{2} K^2
    - \frac{1}{1 + \Theta} \nabla_m a^m \nonumber \\
&\qquad\qquad
    - \frac{F}{2} \left(
          \nabla_{\mathbf{u}} K
        + K_{mn} K^{mn}
        + K^2
      \right)
    - \frac{R}{1 + \Theta}
    \Bigg),
\end{align}
\begin{align}\label{P}
P &= -\frac{\Theta}{3}\left(
        2 \nabla_{\mathbf{u}} K
      + \frac{K^2}{2}
      + \frac{3}{2} K_{mn} K^{mn}
    \right)
    - \frac{\dot{\Theta}}{3}\,K.
\end{align}
These expressions show explicitly how the effective density and pressure are built from the extrinsic
curvature, the acceleration of the preferred congruence, and curvature scalars of $g_{ab}$, together with
the interpolator $\Theta$ and its time derivative.

\paragraph{Heat flux and anisotropic stress.}
The heat flux $q_m$ and anisotropic stress tensor $\pi_{mn}$, which encode departures from a perfect-fluid
form, are given by
\begin{equation}
q_m = \Theta\,a^n K_{mn},
\end{equation}
\begin{align}\label{eq:pi}
\pi_{mn} &= \Theta \Bigg[
      2 t_{(m} K_{n)a} \nabla_{\mathbf{u}} u^a
    - K K_{mn}
    - \nabla_{\mathbf{u}} K_{mn} \nonumber \\
&\qquad\qquad
    + \frac{h_{mn}}{3}\bigl( \nabla_{\mathbf{u}} K + K^2 \bigr)
    \Bigg]
    + \frac{\dot{\Theta}}{6}\left(
          K h_{mn}
        - 3 K_{mn}
      \right).
\end{align}
In the perfectly isotropic, geodesic case ($a_m=0$ and $K_{mn} = K h_{mn}/3$), both $q_m$ and $\pi_{mn}$
vanish, and the effective source reduces to a perfect fluid. Small deviations from geodesy and isotropy
produce correspondingly small but finite $q_m$ and $\pi_{mn}$.

The above expressions provide a detailed account of how the effective energy–momentum tensor arises across
the signature change. The behavior of its components, particularly near the hypersurface $\Sigma_0$ where
$\Theta=-1$ and $\dot\Theta$ can be large, highlights the tight interplay between the local geometry and
the physical quantities induced by the transition. In the next section we quantify these effects in a
near-FRW setting, showing that small heat flux and anisotropic stress leave the geometric inflationary
window largely intact, modifying its duration only by subleading corrections.

\section{Mathematical Treatment for Inflation}\label{Inflation}

In this section we analyse inflation driven by the effective stress tensor
$T_{ab}$ in the presence of small but non-zero heat flux
$q_a$ and anisotropic stress $\pi_{ab}$. Our strategy is to work in a controlled
expansion near the signature--change hypersurface $\Sigma_0$ where $\Theta=-1$,
and to isolate the leading geometric contributions to the effective energy
density, pressure and equation of state.

\subsection{Controlled expansion near the transition}

Near $\Sigma_0$ it is convenient to introduce the small parameter
\begin{equation}
\varepsilon(t)\equiv1+\Theta(t),
\qquad
\varepsilon(0)=0^{+},\qquad 0<\varepsilon\ll1,
\end{equation}
so that $\Theta=-1+\varepsilon$ on the early Lorentzian side. The function
$F(\Theta)$ behaves as
\begin{equation}
F = \frac{\Theta}{1+\Theta}
= \frac{-1+\varepsilon}{\varepsilon}
= -\frac{1}{\varepsilon} + 1,
\end{equation}
and is therefore dominated by $F\simeq -1/\varepsilon$ as
$\varepsilon\to0^{+}$.

In what follows we work in a near--FRW, near--geodesic regime, as made precise in
App.~\ref{app:small-anisotropy}. Concretely, we introduce two small, dimensionless
parameters $(\sigma_K,\sigma_a)$ controlling, respectively, the traceless part of
the extrinsic curvature and the acceleration of the preferred congruence $u^a$,
and assume that $\Theta$, $K_{ab}$, $a^a$ and their first derivatives
along $u^a$ remain finite as $\varepsilon\to0^+$, in accordance with the regular
signature--change conditions.

Under these assumptions, the heat flux $q_a$ and
anisotropic stress $\pi_{ab}$  satisfy
\begin{equation}
q_m = \mathcal{O}(\sigma_a), \qquad \pi_{mn} = \mathcal{O}(\sigma_K+\sigma_a)
\end{equation}
as shown in App.~\ref{app:small-anisotropy}. Thus, in the double-scaling regime
$\sigma\ll1$ and $\varepsilon\ll1$,
the isotropic sector $(\rho,P)$, treating $q_a$ and $\pi_{ab}$ as subleading corrections.

\subsection{Effective energy density and pressure}

Starting from Eqs.~\eqref{eq:rho}--\eqref{P}, the effective energy density can
be recast in the compact form by use of Gauss Codazzi equations
\begin{equation}
\label{density}
\rho = \frac{F}{2}
       \left( \nabla_m a^m -{}^{(3)}\!R
                     + \Theta\,\nabla_{\textbf{u}} K\right),
\end{equation}
while the effective pressure is
\begin{equation}
\label{pressure}
P = -\frac{\Theta}{3}
      \left( 2 \nabla_{\textbf{u}} K
            + \frac{K^2}{2}
            + \frac{3}{2} K_{mn} K^{mn} \right)
    -\frac{\dot{\Theta}}{3} K.
\end{equation}
Near $\Sigma_0$ we have $\Theta=-1+\varepsilon\simeq-1$ and
$F\simeq -1/\varepsilon$. For a nearly geodesic congruence
($\nabla_m a^m=\mathcal O(\varepsilon^\alpha)$) the $\nabla_m a^m$ term in
Eq.~\eqref{density} is subleading, and the potentially divergent pieces arise
from the product $F\Theta\nabla_{\mathbf u}K$ and from the spatial curvature
${}^{(3)}\!R$. Collecting the leading contributions one finds
\begin{align}
\rho &\;\sim\; \frac{1}{2\varepsilon}
               \bigl(\nabla_{\textbf{u}} K + {}^{(3)}\!R\bigr)
               + \mathcal O(\varepsilon^0),
\label{eq:rhoP_divergent-rho}\\
P &\;\sim\; -\frac{\dot \Theta}{3}\,K
             + \mathcal O(\varepsilon^0).
\label{eq:rhoP_divergent-P}
\end{align}
 These expressions capture the
dominant behaviour of the effective fluid near the signature--change surface.

\subsection{Inflation from fast signature change: local geometric criterion}
\label{subsec:modelIndependent}
\begin{figure}[h]
    \centering
    \includegraphics[height=6.5cm,width=0.475\textwidth]{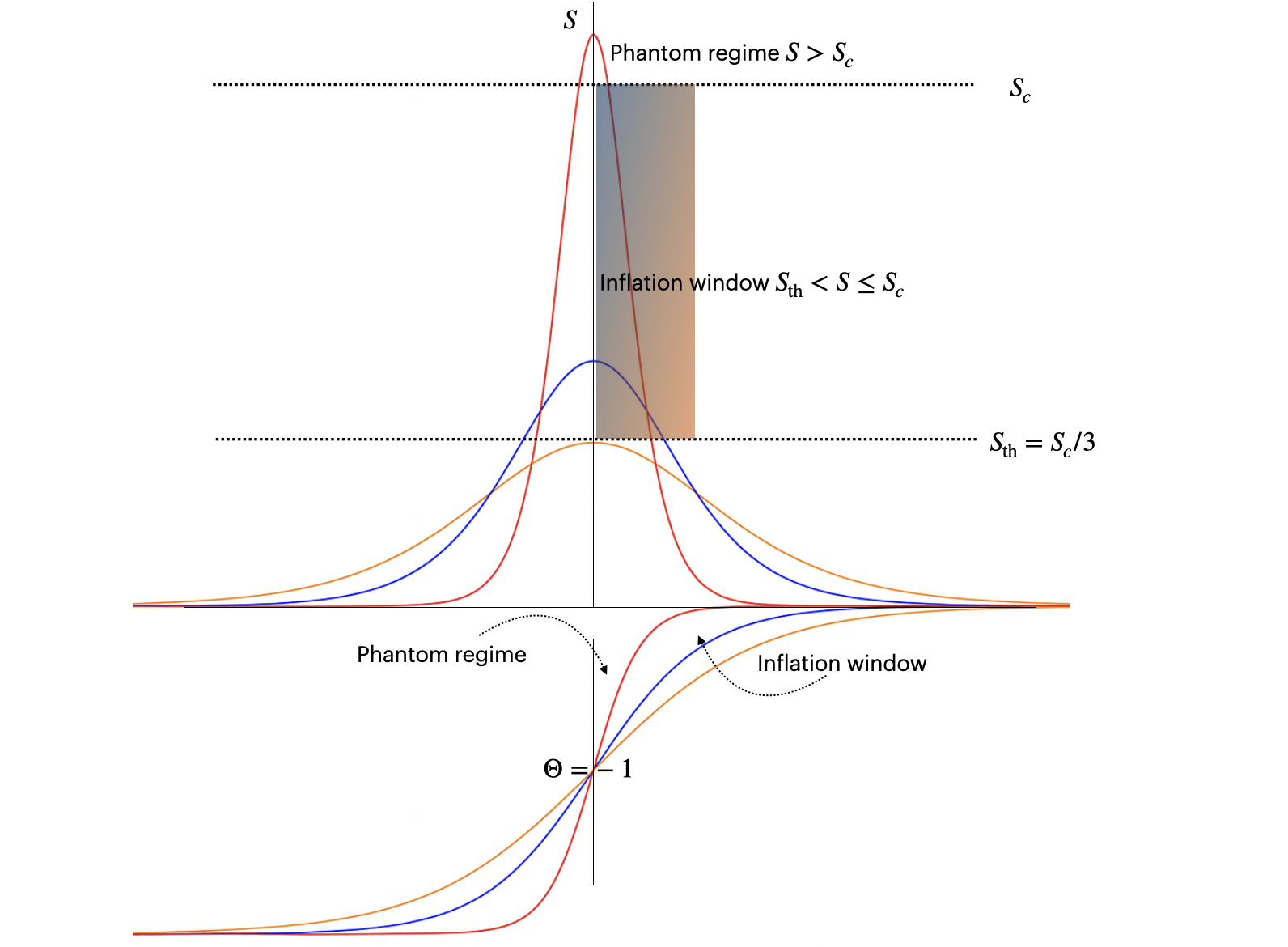}
    \caption{Slope of the interpolator $S(t)=\dot\Theta(\lambda t)$(upper half) for
$\Theta(\lambda t)=\tanh(\lambda t)-1$ (lower half) and three choices of peak slope
$S_0=\lambda$. The horizontal lines show the geometric threshold
$S_{\rm th}=S_c/3$ (corresponding to $w=-1/3$) and the ceiling $S_c$
(corresponding to $w=-1$), with the shaded band marking the non-phantom
inflationary window $S_{\rm th}<S(t)\le S_c$. The orange curve has
$S_0<S_{\rm th}$ and never inflates, the blue curve has
$S_{\rm th}<S_0<S_c$ and realizes geometric inflation with
$-1<w(t)<-1/3$, and the red curve has $S_0>S_c$ and probes a transient
phantom regime $w<-1$ near $t\simeq 0^+$. For the tanh profile, varying
$\lambda$ simply scans the ratio $S_0/S_c$ at fixed background geometry
$(K,{}^{(3)}R)$.}
    \label{fig: Inflation regime}
\end{figure}

We now rewrite the leading expressions in a form adapted to the inflationary
analysis. On the Lorentzian side
$\varepsilon>0$ and we denote
\begin{equation}
S(t)\;\equiv\;\dot\Theta(t),
\qquad
\mathcal S \equiv S(0^{+})>0.
\label{eq:defSlope}
\end{equation}
For the smooth interpolators studied here the microscopic transition width
(the time over which $\Theta$ changes by ${\cal O}(1)$ across $\Sigma_0$)
scales as $\Delta t_{\rm tr}\sim \mathcal S^{-1}$.

\medskip
\noindent\textit{Effective fluid near the transition.}
In the nearly FRW regime with small heat flux $q_a$ and anisotropic stress
$\pi_{ab}$, Eqs.~\eqref{eq:rhoP_divergent-rho}--\eqref{eq:rhoP_divergent-P}
become
\begin{equation}
\rho \;\simeq\; \frac{1}{2\varepsilon}
                  \left(\nabla_{u}K+{}^{(3)}\!R\right),
\qquad
P \;\simeq\; -\frac{\dot\Theta}{3}\,K,
\label{eq:rhoPgeneric}
\end{equation}
It is convenient to define
\begin{equation}
\Delta \;\equiv\; \nabla_{u}K + {}^{(3)}\!R,
\label{eq:DeltaDef}
\end{equation}
so that
\begin{equation}
\rho \;\simeq\; \frac{\Delta}{2\varepsilon},
\qquad
w \;\equiv\; \frac{P}{\rho}
          \;=\; -\,\frac{2\varepsilon\,S\,K}{3\Delta}.
\label{eq:wExactLocal}
\end{equation}
Positivity of the effective density requires
\begin{equation}
\Delta>0.
\label{eq:rhoPos}
\end{equation}

For a closed FRW background with $K=3H$ and ${}^{(3)}\!R=6/a^2$, one has
\begin{equation}
\Delta \;=\; 3\dot H + \frac{6}{a^2},
\label{eq:DeltaFRW}
\end{equation}
which is automatically positive sufficiently close to the no--boundary cap
(where $a$ is small and the curvature term dominates over $3\dot H$). In this
curvature--dominated regime the effective density
$\rho\simeq\Delta/(2\varepsilon)$ is positive even if $\dot H<0$. 

\medskip
\noindent\textit{Acceleration and slope thresholds.}
On the expanding branch ($K>0$) and for $\Delta>0$,
Eq.~\eqref{eq:wExactLocal} can be written as
\begin{equation}
w(t) \;=\; -\,\frac{S(t)}{S_c},
\qquad
S_c \;\equiv\; \frac{3\Delta}{2\,\varepsilon\,K},
\label{eq:ScDef}
\end{equation}
where $S_c>0$ is a purely geometric scale determined by $(K,{}^{(3)}\!R)$ in a
neighbourhood of $\Sigma_0$. Accelerated expansion requires $w<-1/3$, which
is equivalent to
\begin{equation}
w<-\frac{1}{3}
\quad\Longleftrightarrow\quad
S(t) > S_{\rm th},
\qquad
S_{\rm th} \;\equiv\; \frac{S_c}{3}
                     \;=\; \frac{\Delta}{2\,\varepsilon\,K}.
\label{eq:Sthreshold}
\end{equation}
The upper bound for a non--phantom fluid, $w\ge-1$, corresponds to
$S(t)\le S_c$. Hence the non--phantom inflationary window is
\begin{equation}
-1 \,\le\, w(t)<-\frac{1}{3}
\quad\Longleftrightarrow\quad
S_{\rm th} < S(t) \le S_c,
\label{eq:nonPhantomWindow}
\end{equation}
provided $\Delta>0$. Once the Lorentzian branch has been entered, inflation
is driven purely by geometry whenever the interpolator slope $S(t)$ lies
between the geometric threshold $S_{\rm th}$ and the geometric ceiling $S_c$.

\medskip
\noindent\textit{Closed FRW and positivity of $\Delta$.}
In a closed FRW geometry with scale factor $a(t)$ and Hubble parameter
$H\equiv\dot a/a$, one has $K=3H$ and ${}^{(3)}\!R=6/a^2$, so that
\begin{equation}
\Delta \;=\; 3\dot H + \frac{6}{a^{2}}.
\label{eq:DeltaFRW-again}
\end{equation}
It is useful to collect the departure from exact de Sitter expansion into the
dimensionless parameter
\begin{equation}
\epsilon_G \;\equiv\; \frac{\Delta}{3H^2}
                   \;=\; \frac{\dot H}{H^2} + \frac{2}{a^2 H^2},
\qquad
\Delta = 3\epsilon_G H^2.
\label{eq:epsilonGDef}
\end{equation}
Introducing the kinematic Hubble-flow parameter
$\epsilon_H\equiv-\dot H/H^2\ge0$ and the comoving Hubble radius
$r_H\equiv (aH)^{-1}$, this can be written as
\begin{equation}
\epsilon_G = -\epsilon_H + \frac{2}{a^2 H^2}
           = -\epsilon_H + 2 r_H^2
           = -\epsilon_H + 2|1-\Omega|.
\end{equation}
The condition $\Delta>0$ required for a positive effective density
$\rho\simeq\Delta/(2\varepsilon)$ is therefore equivalent to
\begin{equation}
\epsilon_G>0
\quad\Longleftrightarrow\quad
2 r_H^2 > \epsilon_H,
\end{equation}
i.e.\ the curvature contribution $2/(a^2H^2)$ (or, equivalently, the squared
comoving Hubble radius) dominates over the negative term $-\dot H/H^2$.
Near the no-boundary cap one has $aH\ll1$, so $r_H^2\gg1$ and the inequality
is easily satisfied even when $\epsilon_H$ is small and positive. In contrast,
in a spatially flat slow-roll background with ${}^{(3)}\!R=0$ one finds
$\Delta=3\dot H=-3\epsilon_H H^2\le0$, so the geometric fluid associated with
$\widehat g_{ab}$ would carry non-positive energy density and cannot by itself
drive inflation. Our mechanism is therefore intrinsically tied to the
curvature-dominated, closed initial geometry suggested by the no-boundary
proposal. When $\Delta\to0$ the effective density $\rho\to0$ and the
fluid description ceases to be useful; in that limit the signature-change
source simply switches off rather than describing a physical fluid with
divergent $w$.

A convenient way to visualize the constraint $\Delta>0$ is
\begin{equation}
\frac{6}{a^{2}}
    > -3\dot H
    \;\Longleftrightarrow\;
    \frac{1}{a^{2}}
        > \frac{1}{2}\,\epsilon_H H^{2},
\qquad
\epsilon_H \equiv -\frac{\dot H}{H^{2}} \ge 0 .
\label{eq:DeltaPosFRW}
\end{equation}
Defining $a_\star$ as the scale factor on some early Lorentzian slice just
after the transition, and writing $a(t)=a_\star e^{N(t)}$ with $N(t)$ the
number of e-folds since that slice, Eq.~\eqref{eq:DeltaPosFRW} implies
\begin{equation}
\begin{aligned}
a(t)^{2}
    &< \frac{2}{\epsilon_H H^{2}}
       \;\Rightarrow\;
       N(t) < N_{\max}, \\[6pt]
N_{\max}
    &\equiv \frac{1}{2}
       \ln\!\left[\frac{2}{\epsilon_H H^{2} a_\star^{2}}\right].
\end{aligned}
\label{eq:NmaxDef}
\end{equation}

where we have treated $H$ and $\epsilon_H$ as slowly varying over the short
geometric window. Since $H a_\star$ is the ratio of the initial curvature
radius to the Hubble radius, taking $H a_\star\ll 1$ (as is natural for a
small no-boundary 3-sphere) makes $N_{\max}$ parametrically large.

\medskip
\noindent\textit{Graceful exit from geometric inflation.}
For the tanh, generalized logistic, and power--law interpolators considered in
App.~\ref{app:interpolators}, the slope $S(t)$ has a single non-degenerate
maximum at $t\simeq 0^{+}$ and then decreases monotonically for $t>0$ ( depicted for tanh case in fig \ref{fig: Inflation regime}).
On the relevant interval, the evolution of
$w(t)$ is governed entirely by $S(t)$ via Eq.~\eqref{eq:ScDef}:
\begin{itemize}[leftmargin=*]
\item if $S_0\equiv S(0^{+})\le S_{\rm th}$, the interpolator never crosses
      the acceleration threshold and no inflation occurs;
\item if $S_{\rm th}<S_0\le S_c$, the interpolator enters the non--phantom
      window~\eqref{eq:nonPhantomWindow} and drives a phase with
      $-1\le w(t)<-1/3$;
\item as $S(t)$ decays and first hits $S(t_{\rm end})=S_{\rm th}$, the
      effective equation of state reaches $w=-1/3$ and inflation ends.
\end{itemize}
Thus both the onset and the exit of geometric inflation are determined
locally: the onset requires a sufficiently steep rise of $\Theta$ immediately
after the transition, and the exit is automatic once the interpolator slope
falls below the geometric threshold $S_{\rm th}$.

\begin{figure*}[t]
    \centering
    \includegraphics[width=0.95\textwidth]{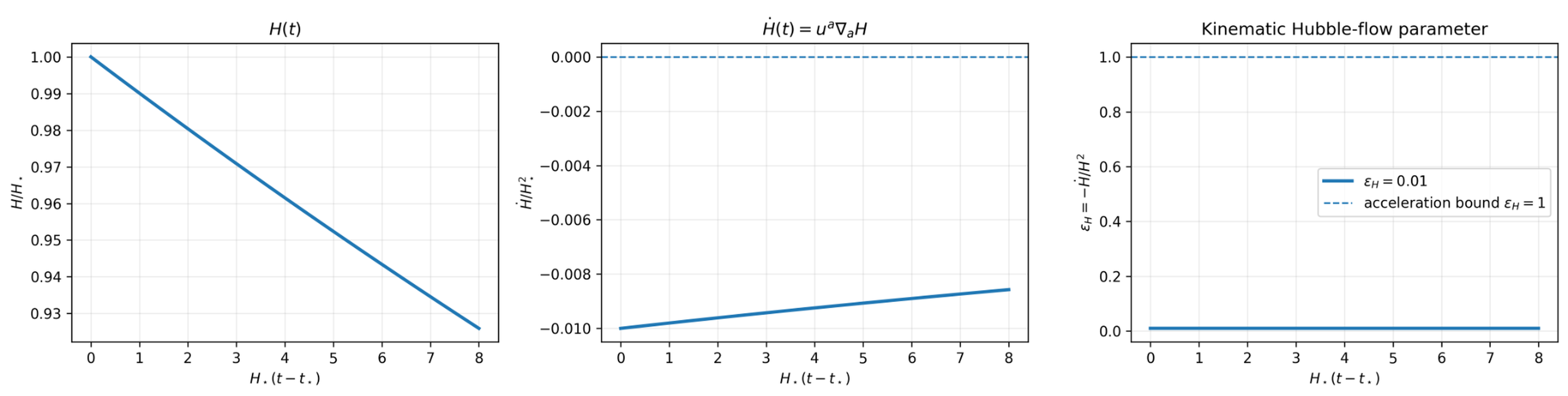}
    \caption{Lorentzian expansion diagnostics:-
The plotted variables are the local geometric Hubble scalar \(H=K/3\),
its derivative \(\dot H=u^a\nabla_aH\), and the kinematic quantity
\(\epsilon_H=-\dot H/H^2\). The representative curves use
\(\epsilon_H=10^{-2}\),
\(H/H_\star=[1+\epsilon_HH_\star(t-t_\star)]^{-1}\) and
\(\dot H/H_\star^2=-\epsilon_H[1+\epsilon_HH_\star(t-t_\star)]^{-2}\).}
    \label{fig:Hdiag}
\end{figure*}

\medskip
\noindent\textit{Local Hubble scalar and its derivative.} To make the Lorentzian expansion variables explicit, we plot the local Hubble scalar and its derivative in Fig.~\ref{fig:Hdiag}. These quantities are purely geometric here and can be written as: $H\equiv K/3$ and $\dot H\equiv u^a\nabla_aH=\nabla_uK/3$. In the exact FLRW limit this reduces to the usual $H=\dot a/a$, while in the near-FRW case it remains the local volume-expansion scalar. These variables are precisely the ones entering the geometric effective source. Indeed, $\Delta=\nabla_uK+{}^{(3)}R=3\dot H+{}^{(3)}R$, and near the transition $\rho\simeq\Delta/(2\varepsilon)$ and $P\simeq-\dot\Theta K/3=-S(t)H$. Consequently, $S_c(t)=3\Delta/(2\varepsilon K)=\Delta/(2\varepsilon H)$, $S_{\rm th}(t)=S_c(t)/3$, and $w(t)=-S(t)/S_c(t)$. Thus the interpolator does not define $H$; rather, it determines when the Lorentzian geometric branch satisfies $S_{\rm th}(t)<S(t)\le S_c(t)$.

For visualization, Fig.~\ref{fig:Hdiag} shows a representative Lorentzian branch. We choose a simple ordinary branch with $H>0$ and $\dot H<0$, so the plot is a diagnostic of the Lorentzian expansion variables appearing in the geometric source. It is not meant as a unique solution of the full post-transition dynamics. For the representative curves we use the kinematic quantity $\epsilon_H\equiv-\dot H/H^2=10^{-2}$, giving $H/H_\star=[1+\epsilon_HH_\star(t-t_\star)]^{-1}$ and $\dot H/H_\star^2=-\epsilon_H[1+\epsilon_HH_\star(t-t_\star)]^{-2}$. The de Sitter limit is recovered for $\epsilon_H\to0$.

\medskip
\noindent\textit{Phenomenological e-fold estimate from the geometric window.}
The number of e--folds accumulated during the geometric inflationary phase is
\begin{equation}
N_e \;=\; \ln\frac{a(t_{\rm end})}{a(t_\star)}
      \;=\; \int_{t_\star}^{t_{\rm end}} H(t)\,dt,
\label{eq:Ne-def}
\end{equation}
where $t_\star$ is a time just after the transition with
$S(t_\star)\lesssim S_0$ and $t_{\rm end}$ is fixed by
$S(t_{\rm end})=S_{\rm th}$. This relation is exact and does not assume a
constant equation of state or a specific matter model.

To obtain analytic estimates we now make an additional, purely kinematic
assumption about the background during the geometric window: we treat it as
quasi--de Sitter, in the sense that the Hubble rate varies slowly on
$[t_\star,t_{\rm end}]$. Writing $H_\star\equiv H(t_\star)$ and
$H(t)=H_\star+\delta H(t)$, with $|\dot H|/H_\star^2$ small on the interval,
one finds (see App.~\ref{app:efolds} for details) that
\begin{equation}
N_e \;=\; H_\star \Delta t_{\rm inf}
        \;+\;\mathcal O\!\left(\epsilon_{\rm eff}\,H_\star^2 \Delta t_{\rm inf}^2\right),
\qquad
\Delta t_{\rm inf}\equiv t_{\rm end}-t_\star,
\label{eq:Ne-approx}
\end{equation}
where $\epsilon_{\rm eff}$ is an effective slow--variation parameter on the
interval. Using $N_e\simeq H_\star \Delta t_{\rm inf}$ in the leading term,
the relative correction is parametrically of order
$\delta N_e/N_e\sim\epsilon_{\rm eff}N_e$, so that for
$\epsilon_{\rm eff}N_e\ll1$ (e.g.\ $\epsilon_{\rm eff}\lesssim10^{-3}$ when
$N_e\sim 50$--$60$) the approximation $N_e\simeq H_\star\Delta t_{\rm inf}$
is accurate at the few-percent level. This estimate does not rely on a
constant equation of state; it follows solely from the quasi--de Sitter
assumption that $H(t)$ varies slowly on the geometric window.

The remaining input is $\Delta t_{\rm inf}$, which is fixed by the
interpolator shape and the geometric scales $S_c$ and $S_{\rm th}$. Close to
$\Sigma_0$ we may treat $S_c$ as approximately constant and equal to its
value on an early Lorentzian slice,
$S_c\simeq S_c(0^{+})=3\Delta/(2\varepsilon K)$, and solve
$S(t_{\rm end})=S_{\rm th}=S_c/3$ for the interpolators of
App.~\ref{app:interpolators}. The corresponding profiles, initial slopes, and inflationary durations in this approximately constant-threshold limit are summarized in Table \ref{tab:interpolators-clean}. For the tanh profile, one finds
\begin{equation}
\Delta t_{\rm inf}^{(\tanh)}
    \;=\; \frac{1}{S_c}\,\mathrm{arcosh}\!\sqrt{3}
    \;\simeq\; \frac{2\varepsilon}{3\,\epsilon_G H_\star}\,
               \mathrm{arcosh}\!\sqrt{3},
\end{equation}
while the normalized power--law (arctan) profile yields
\begin{equation}
\Delta t_{\rm inf}^{(\mathrm{PL})}
    \;=\; \frac{2\sqrt{2}}{\pi S_c}
    \;\simeq\; \frac{4\sqrt{2}\,\varepsilon}{3\pi\,\epsilon_G H_\star},
\end{equation}
where in the last step we used $\Delta=3\epsilon_G H_\star^2$ and
$K=3H_\star$ on the reference slice. Inserting these into
Eq.~\eqref{eq:Ne-approx} and retaining the leading quasi--de Sitter term gives
the scaling
\begin{equation}
\begin{aligned}
N_e^{(\tanh)}
    &\simeq H_\star\,\Delta t_{\rm inf}^{(\tanh)}
     \;\approx\; 0.76\,\frac{\varepsilon}{\epsilon_G}, \\[6pt]
N_e^{(\mathrm{PL})}
    &\simeq H_\star\,\Delta t_{\rm inf}^{(\mathrm{PL})}
     \;\approx\; 0.60\,\frac{\varepsilon}{\epsilon_G},
\end{aligned}
\end{equation}
up to relative corrections $\sim\epsilon_{\rm eff}N_e$. These expressions should be read as frozen-threshold analytic expressions: they are obtained by holding the geometric threshold $S_{\rm th}=\epsilon_GH/(2\varepsilon)$ approximately fixed while solving for the crossing of the interpolator slope. This limit is useful for comparing interpolator profiles, but it is not the fully dynamical e--fold prediction in a curvature-supported branch. When the closed-curvature term in $\Delta=3\dot H+6/a^2$ is important, ${\cal K}=1/(a^2H^2)$ is diluted during expansion and hence $S_{\rm th}$ and $S_c$ move. The relevant long-duration condition is therefore the persistence of the moving band $S_{\rm th}(N)<S(N)\le S_c(N)$, as discussed in App.~\ref{app:efolds}.

And this is the reason that at first sight, the expression scaling $N_e\propto\varepsilon/\epsilon_G$ may appear to be in tension with a strongly curvature-dominated initial slice. If ${\cal K}_\star\equiv1/(a_\star^2H_\star^2)\gg1$ and $\epsilon_H<<1$, then $\epsilon_{G,\star}=-\epsilon_{H,\star}+2{\cal K}_\star\simeq2{\cal K}_\star$, so a strictly frozen-threshold substitution would suggest a small number of e--folds. This conclusion is an artifact of freezing the threshold. For a branch with approximately constant $\epsilon_H=\beta$, ${\cal K}(N)={\cal K}_\star e^{-2(1-\beta)N}$ and $\epsilon_G(N)=2{\cal K}_\star e^{-2(1-\beta)N}-\beta$. Thus large initial curvature shifts the moving thresholds rather than simply imposing $N_e\propto1/{\cal K}_\star$. In the moving-threshold description the duration is controlled by the interval over which ${\cal R}(N)\equiv S(N)/S_{\rm th}(N)$ remains in the non--phantom band $1<{\cal R}(N)\le3$. Large initial curvature therefore does not algebraically force a small $N_e$; the relevant question is whether the moving band persists for sufficiently long.

\section{Conclusion and Discussion}
\label{Conclusion}

We have investigated a purely geometric mechanism for early--universe inflation based on a smooth change of metric signature. The construction is defined by a covariant continuation
\(
\widehat g^{ab} = g^{ab} - \Theta\,u^a u^b
\)
along a preferred timelike congruence $u^a$, with a transition hypersurface $\Sigma_0$ at $\Theta=-1$ on which the continued metric is degenerate but curvature invariants remain finite. Interpreting
\(
T_{ab} \equiv \widehat G_{ab} - G_{ab}
\)
as an effective source for the Lorentzian metric $g_{ab}$ yields a localized, purely geometrical stress tensor supported in a thin neighborhood of $\Sigma_0$.

In the nearly FRW regime with small heat flux and anisotropic stress, the effective fluid just on the Lorentzian side is characterized by
\(
\rho \simeq \Delta/(2\varepsilon)
\)
and
\(
P \simeq -\dot\Theta\,K/3
\),
with $\varepsilon \equiv 1+\Theta$ and
\(
\Delta \equiv \nabla_u K + {}^{(3)}\!R
\).
For the (almost) comoving congruence we adopt, $K>0$ characterizes the expanding branch, and for $\Delta>0$ the local equation of state takes the simple form
\(
w(t)=P/\rho=-S(t)/S_c
\),
where $S(t)=\dot\Theta(t)$ and
\(
S_c = 3\Delta/(2\,\varepsilon K)
\)
is a purely geometric scale constructed from $(K,{}^{(3)}\!R)$ in a neighborhood of $\Sigma_0$. Accelerated expansion requires $w<-1/3$ and hence
\(
S_{\rm th}<S(t)\le S_c
\)
with $S_{\rm th} \equiv S_c/3$, while the non–phantom bound corresponds to $w\ge -1$. Thus the onset and exit of inflation are controlled entirely by the local slope of the interpolator relative to the geometric scales $S_{\rm th}$ and $S_c$, without introducing an inflaton sector. We also clarify the behavior of the effective-fluid variables near the
degenerate transition surface. The Lorentzian effective-fluid description
should not be extrapolated all the way to
\(\varepsilon=1+\Theta=0\). Near the Lorentzian side one has
\(\rho\simeq \Delta/(2\varepsilon)\) and
\(S_{\rm th}\simeq \Delta/(2\varepsilon K)\); hence both quantities diverge
formally as \(\varepsilon\to0^+\), while \(\Delta\) and \(K\) may remain
finite. The divergence of \(S_{\rm th}\) is not an independent geometric
effect, but follows from the same effective-fluid normalization, since
\(S_{\rm th}=\rho/K\). Thus the apparent divergence is not a curvature
singularity of the continued geometry, but signals the breakdown of the
Lorentzian fluid parametrization exactly at the degenerate hypersurface.

\begin{figure}[t]
    \centering
    \includegraphics[width=0.48\textwidth]{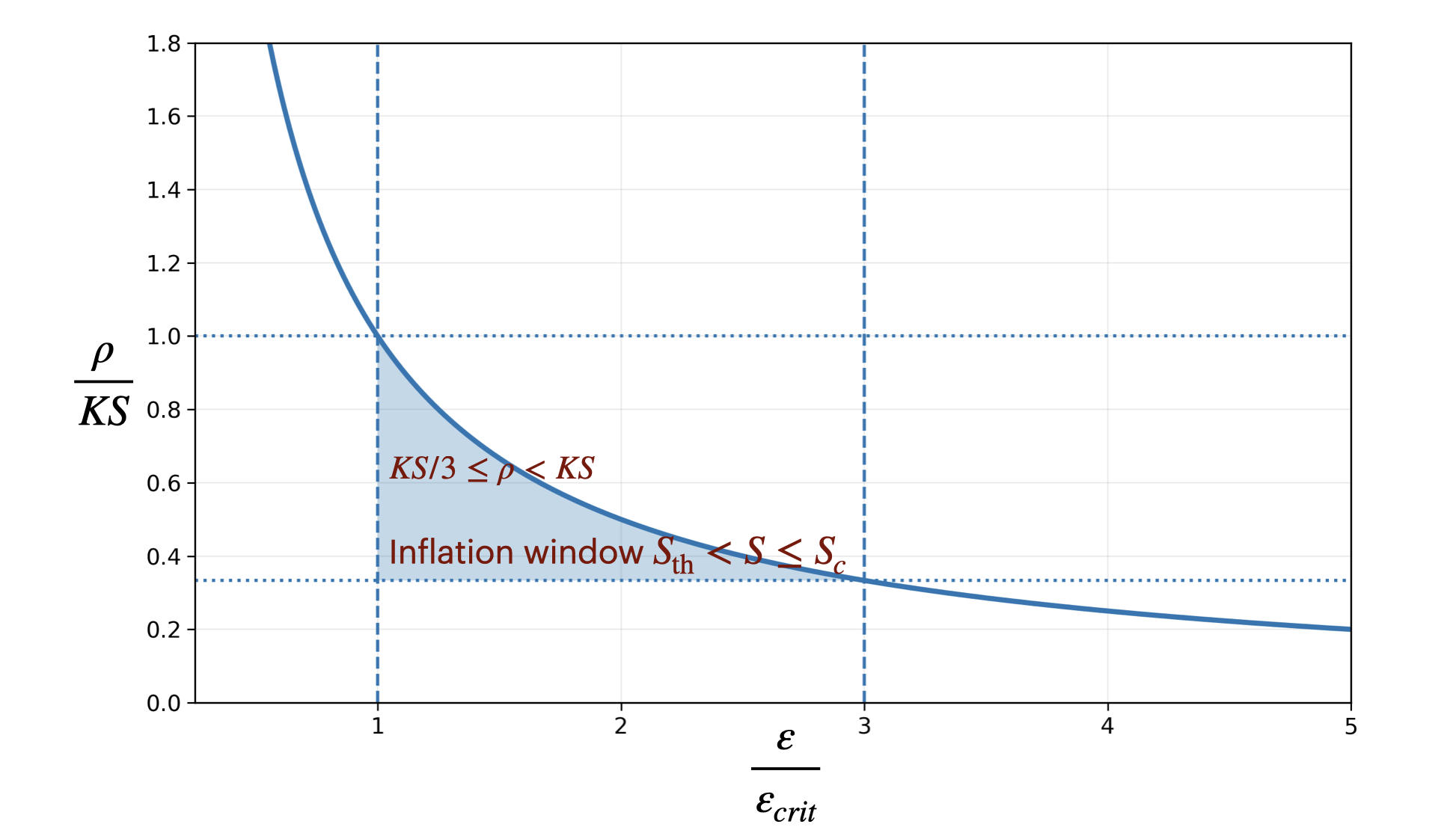}
    \caption{
Finite-\(\varepsilon\) density window for the geometric effective fluid.
Defining
\(\varepsilon_{\rm crit}\equiv\Delta/(2KS)\), the near-transition density
\(\rho\simeq\Delta/(2\varepsilon)\) gives
\(\rho/(KS)=\varepsilon_{\rm crit}/\varepsilon\). The curve plots this
dimensionless relation. The shaded region
\(1<\varepsilon/\varepsilon_{\rm crit}\le3\) is equivalent to the
non-phantom inflationary density range \(KS/3\le\rho<KS\), or
\(S_c/3<S(t)\le S_c\). Hence the allowed inflationary regime occurs at finite
\(\varepsilon>0\), away from the transition hypersurface.
}
    \label{fig:rhoepsilon}
\end{figure}

The effective description should instead be initialized on a first regular
Lorentzian slice \(\Sigma_\star\) with
\(\varepsilon_\star\equiv\varepsilon_{\min}>0\). On this slice,
\(\rho_{\star}\simeq \Delta_\star/(2\varepsilon_{\min})\) and
\(S_{{\rm th},\star}\simeq
\Delta_\star/(2\varepsilon_{\min}K_\star)\) are finite, and inflation begins
only if \(S(t_\star)>S_{{\rm th},\star}\). Equivalently, for finite
\(S(t)\), the accelerated phase starts only after
\(\varepsilon>\Delta/(2KS)\). Using
\(S_c=3\Delta/(2\varepsilon K)=3\rho/K\), the non-phantom window
\(S_c/3<S(t)\le S_c\) implies
\(KS(t)/3\le\rho<KS(t)\).  This finite-density window is illustrated in
Fig.~(\ref{fig:rhoepsilon}). Thus, for finite \(K\) and finite interpolator
slope, the effective density in the allowed inflationary interval is finite
and positive . The apparent divergence therefore lies outside the
finite-slope Lorentzian inflationary regime and marks only the boundary of
validity of the effective-fluid description.

\begin{figure}[t]
    \centering
    \includegraphics[width=0.72\linewidth]{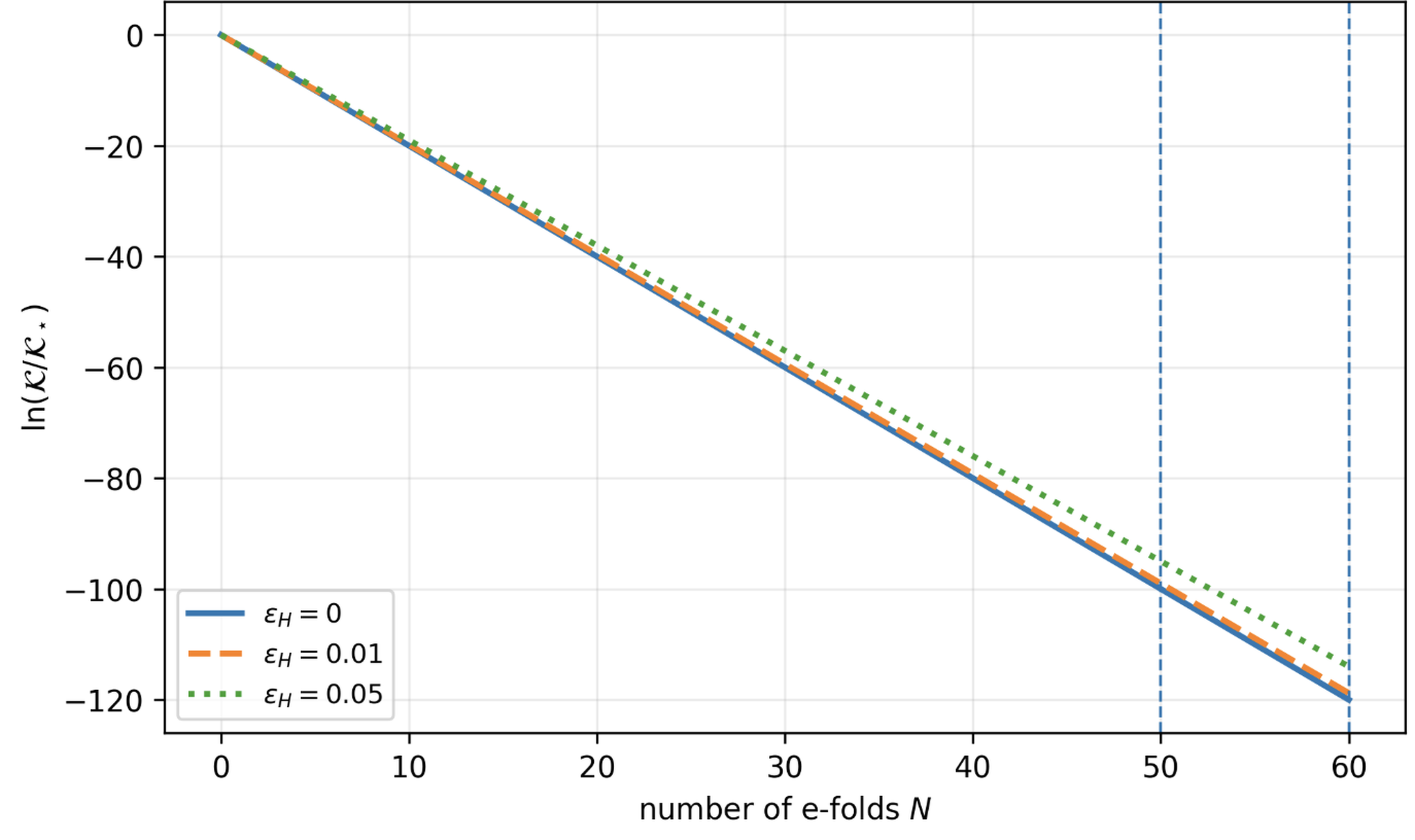}
    \caption{
    Dilution of the dimensionless curvature fraction ${\cal K}=1/(a^2H^2)=|\Omega_k|$ during accelerated expansion. For constant $\epsilon_H$, ${\cal K}(N)/{\cal K}_\star=e^{-2(1-\epsilon_H)N}$.The vertical dashed lines indicate $N=50$ and $N=60$. Thus even if the closed curvature is important near the transition, it is exponentially suppressed after $50$--$60$ e-folds.
    }
    \label{fig:curvature-dilution}
\end{figure}

A key feature of this mechanism is that it relies on curvature domination rather than slow roll. In a closed FRW background,
\(
K=3H
\)
and
\(
{}^{(3)}\!R=6/a^2
\)
imply
\(
\Delta = 3\dot H + 6/a^2.
\)
Near the transition, the positive curvature contribution $6/a^2$ can dominate over $3\dot H=-3\epsilon_H H^2\le0$, ensuring $\Delta>0$ and hence $\rho>0$ even when $H$ varies slowly. In terms of the geometric parameter
\(
\epsilon_G = \Delta/(3H^2) = \dot H/H^2 + 2/(a^2 H^2),
\)
the curvature term $2/(a^2 H^2)$ is proportional to the inverse squared comoving Hubble radius, so the requirement $\epsilon_G>0$ encapsulates the fact that the initial closed geometry shrinks the comoving Hubble scale and drives a curvature–dominated inflationary phase. By contrast, the usual flat slow–roll limit with ${}^{(3)}\!R=0$ and $\dot H<0$ would give $\Delta<0$ and an unphysical effective fluid. This curvature-supported initial condition should not be confused with a late-time prediction of large spatial curvature. Defining the dimensionless curvature fraction \({\cal K}\equiv 1/(a^2H^2)=|\Omega_k|\), one has \(d\ln{\cal K}/dN=-2(1-\epsilon_H)\)  with $N=\ln(a/a_\star)$. Hence, during accelerated expansion, for which \(\epsilon_H<1\), \({\cal K}\) decreases monotonically; in the quasi--de Sitter limit this gives \({\cal K}_{\rm end}\simeq {\cal K}_\star e^{-2N_e^{\rm(geom)}}\). Figure~(\ref{fig:curvature-dilution}) illustrates the exponential dilution over $N=50$--$60$ e-folds. The same closed curvature that supports the geometric source near the transition is therefore diluted during the accelerated phase, allowing the late Lorentzian universe to be observationally close to spatial flatness. Our construction consequently singles out closed, curvature-dominated initial data, consistent with the compact $S^3$ spatial slices that typically arise in no-boundary-type cosmological models~\cite{HartleHawking:1983,GibbonsHartle:1990}.

The same positivity condition $\Delta>0$ also constrains the duration over which the curvature-supported effective density description is applicable. Writing $a(t)=a_\star e^{N(t)}$ and introducing the kinematic parameter $\epsilon_H=-\dot H/H^2$, one obtains the upper bound $N(t)<N_{\max}$ in Eq.~\eqref{eq:NmaxDef}, beyond which $\Delta$ would change sign. This bound is a necessary consistency condition, but it is not by itself the full e--fold calculation. The closed-form estimates obtained from smooth interpolators should be viewed as frozen-threshold expressions. In the curvature-dominated case the threshold generally moves because ${\cal K}=1/(a^2H^2)$ dilutes during expansion, so the long-duration question is controlled by the persistence of $S_{\rm th}(N)<S(N)\le S_c(N)$, as summarized in App.~\ref{app:efolds}. 

Conceptually, the picture is simple: a regular Euclidean cap provides smooth initial data on $\Sigma_0$, the effective geometric stress tensor generated by the covariant signature change drives a transient phase with $-1\le w<-1/3$ as $\Theta$ rises from $-2$ to $0$, and inflation ends automatically when the interpolator slope drops below the geometric threshold $S_{\rm th}$. The horizon and flatness problems are addressed by this curvature–dominated, inflaton–free phase, whose properties are determined by local geometric quantities rather than by a scalar potential.

Several open directions follow from this work. On the theoretical side, a natural next step is to develop the perturbation theory associated with the geometric source $T_{ab}=\widehat G_{ab}-G_{ab}$, and to compute the corresponding scalar and tensor power spectra, as well as any associated non-Gaussianities or feature-like imprints linked to the finite duration of the signature–change–driven phase. This analysis would clarify how the geometric parameters of our construction—such as $(\epsilon_G,\varepsilon)$ and the interpolator profile $S(t)$—translate into standard observational parameters $(n_s,r,f_{\mathrm{NL}},\ldots)$, and to what extent they are compatible with current CMB and large-scale structure constraints, including bounds on residual spatial curvature.

Beyond pure model building, the covariant formalism developed here provides a controlled, local description of how a regular Euclidean cap can supply effective initial data for the Lorentzian universe, as a stand-alone geometric inflationary phase. Embedding this construction into concrete quantum-cosmological frameworks (for example specific realizations of the no–boundary wave function or other path-integral proposals) would clarify the microscopic origin of the interpolator $\Theta$ and sharpen the link between Euclidean boundary data and late-time observables. Taken together, these directions suggest that covariant signature change is not only a technically consistent way to replace the big-bang singularity, but also a potentially testable geometric mechanism for the earliest inflationary dynamics.

\section{Acknowledgements}
The authors thank the anonymous referee for constructive comments and suggestions, which helped improve the clarity and presentation of the manuscript.
\appendix
\label{app:data1}
\section{Geometry of signature change}
\label{app:geometry}

\subsection{Conventions}

Except when indicated otherwise, we work in units with $c = 1$ and $\hbar = 1$.
Lorentzian metric signature is $(-,+,+,+)$ and Euclidean metric signature is
$(+,+,+,+)$. The projection tensor onto the spatial hypersurfaces orthogonal to
the preferred congruence $u^a$ is
\begin{equation}
h_{ab} \equiv g_{ab} + u_a u_b,
\end{equation}
and our convention for the extrinsic curvature of these hypersurfaces is
\begin{equation}
K_{ab} = \nabla_a u_b + u_a a_b,
\qquad
a^a \equiv u^b \nabla_b u^a .
\end{equation}
Throughout this appendix, all unhatted quantities (e.g.\ $R_{ab}{}^{cd}$,
$G^a{}_b$, $K_{ab}$) are associated with the Lorentzian metric $g_{ab}$.

\subsection{Useful formulae}

We summarize here the main geometric identities for the class of covariant continued
metrics $\widehat g_{ab}$ used in the main text (see also Ref.~\cite{My}).
The covariant continuation is defined by
\begin{align}\label{app:def-metric}
\widehat g^{ab} &= g^{ab} - \Theta u^a u^b,
&
\widehat g_{ab} &= g_{ab} + F t_a t_b,
&
t_a &= g_{ab} u^b ,
\end{align}
with
\begin{equation}
F = \frac{\Theta}{1+\Theta},
\qquad
\dot F = \frac{\dot \Theta}{(1+\Theta)^2},
\end{equation}
and
\begin{equation}
\nabla_a F = -\dot F\,t_a,
\qquad
\nabla_a \dot F = -\ddot F\,t_a - \dot F\,a_a .
\end{equation}
Here $u^a$ is the (unit) velocity field of the preferred congruence,
$a^a = u^b \nabla_b u^a$ is its acceleration, and over-dots denote derivatives
along $u^a$, e.g.\ $\dot X \equiv u^a \nabla_a X$.

The Christoffel symbols of $\widehat g_{ab}$ can be written as
\begin{equation}
\widehat \Gamma^a{}_{bc}
  = \Gamma^a{}_{bc} + C^a{}_{bc},
\end{equation}
with
\begin{equation}
C^a{}_{bc}
 = F\bigl[(1-F) u^a K_{(bc)} - a^a t_b t_c\bigr]
   - \frac{\dot \Theta}{2(1+\Theta)}\,t_b t_c u^a .
\label{app:connection}
\end{equation}

\begin{widetext}

The Riemann tensor can be expressed as
\begin{align}
\widehat R_{ab}{}^{cd}
 &= R_{ab}{}^{cd}
  + 2\Theta\left(
    - u^{[c} R_{abm}{}^{d]} u^m
    - K^{d}{}_{[a} K_{b]}{}^{c}
    + 2 t_{[a} a_{b]} a^{[c} u^{d]}
    + 2 u^{[c} (\nabla_{[a} a^{d]}) t_{b]}
    \right)
  + 2\dot \Theta\, u^{[c} K_{[a}{}^{d]} t_{b]} .
\label{app:Riemann}
\end{align}
where $[\cdots]$ denotes antisymmetrization, 
$A_{[ab]} \equiv \tfrac12 (A_{ab} - A_{ba})$.
. The associated Ricci tensor, Ricci scalar and Einstein tensor are
\begin{align}
\widehat R^a{}_{c}
 &= (1+\Theta)\, R^a{}_{c}
    - \Theta\left(
       {}^{(3)}R^a{}_{c}
       - t_c C^a
       + t_c a^b K^a{}_{b}
       - a^a a_c
       - g^{la} h^r{}_l \nabla_r a_c
       + u^a t_c \nabla_b a^b
     \right)
    + \frac{1}{2}\,\dot \Theta \left(\pi^a{}_{c} + K \delta^a{}_{c}\right), 
\label{app:Ricci-mixed}
\\
\widehat R
 &= (1+\Theta) R
    + \Theta\left(-\,{}^{(3)}R + 2\nabla_b a^b\right)
    + \dot \Theta\,K,
\label{app:Ricci-scalar}
\\
\widehat G^a{}_{c}
 &= (1+\Theta)\, G^a{}_{c}
  - \Theta \Bigl(
      {}^{(3)}G^a{}_{c}
      + \tfrac{1}{2}\,{}^{(3)}R\, u^a t_c
      - t_c C^a
      - t_c a^b K^a{}_b
      - a^a a_c
      + u^a t_c \nabla_b a^b
      - g^{la} h^r{}_l \nabla_r a_c
    \Bigr)
  + \frac{1}{2}\,\dot \Theta\, p^a{}_{c}.
\label{app:Einstein}
\end{align}
Here ${}^{(3)}R^a{}_{c}$ and ${}^{(3)}G^a{}_{c}$ are the intrinsic Ricci and
Einstein tensors associated to metric $h_{ab}$, and
\begin{equation}
C^a \equiv D_b K^{ab} - D^a K,
\qquad
p^a{}_{c} \equiv K^a{}_{c} - K h^a{}_{c},
\end{equation}
with $D_a$ covariant derivative associated to the induced metric $h_{ab}$, i.e. $D_a h_{mn}=0$. Finally, the Kretschmann
scalar can be written as,
\begin{align}
\widehat S \equiv \widehat R_{abcd} \widehat R^{abcd}
 &= S
  + \Theta\left(
      8 R_{ab}{}^{cd} u^{[a} \nabla_{[c} K_{d]}{}^{b]}
      - 4 R_{ab}{}^{cd} K_{[c}{}^{b} K_{d]}{}^{a}
    \right)
\nonumber\\
 &\quad
  + 4\Theta^2\Bigl[
      (\nabla_{\mathbf{u}} K_b{}^{d})(\nabla_{\mathbf{u}} K^b{}_{d})
      + 2(\nabla_{\mathbf{u}} K^b{}_{d}) K^{ad} K_{ba}
      + K^{db} K_{cd} K^{ac} K_{ba}
\nonumber\\
 &\qquad\qquad\qquad
      + \tfrac{1}{2}(K_{mn} K^{mn})^2
      - \tfrac{1}{2} K^{cb} K_{ac} K^{da} K_{bd}
    \Bigr]
\nonumber\\
 &\quad
  + 2\Theta \dot \Theta \left(
      K^b{}_{c} \nabla_{\mathbf{u}} K_b{}^{c}
      + K^{ac} K_{ba} K^b{}_{c}
    \right)
  + 2\dot \Theta\, u^{[a} K^{b]}{}_{[c} t_{d]} R_{ab}{}^{cd}
  + \dot \Theta^2 K^{bd} K_{bd}.
\label{app:Kretschmann}
\end{align}

\end{widetext}

We see that $\widehat S$ is a polynomial in $\Theta$, $\dot \Theta$,
$K_{ab}$, $a^a$, their first $u$–derivatives and the background curvature,
with \emph{no factors of $(1+\Theta)^{-1}$}. In particular, if
\begin{itemize}
\item $\Theta$ is at least $C^2$ across $\Sigma_0$ with finite
      $\Theta|_{\Sigma_0}=-1$ and $\dot \Theta|_{\Sigma_0}$,
\item $K_{ab}$, $a^a$ and their first $u$–derivatives are bounded in a
      neighborhood of $\Sigma_0$\cite{Hayward}, and
\item the background curvature invariants of $g_{ab}$ are finite there,
\end{itemize}
then $\widehat S$ remains finite as $\Theta \to -1$. Under these
regularity conditions the hypersurface $\Sigma_0:\Theta=-1$ is a
curvature–regular, codimension–one surface across which the continued metric
$\widehat g_{ab}$ becomes degenerate but does not develop curvature
singularities. This is the sense in which the signature–change hypersurface
is ``degenerate but non–singular'' in the main text.

\section{Small-anisotropy and near-geodesic scaling}
\label{app:small-anisotropy}

In this appendix we justify the near--FRW, near--geodesic assumptions used in
Sec.~\ref{Inflation} by exhibiting a controlled scaling under which the heat flux
$q_a$ and anisotropic stress $\pi_{ab}$ associated with the effective source remain
subleading near the signature--change surface
$\Sigma_0:\Theta=-1$ when compared to the isotropic energy density $\rho$.

Throughout we work on the Lorentzian side $\Theta>-1$ and introduce
\begin{equation}
\varepsilon(t) \;\equiv\; 1+\Theta(t),
\qquad \varepsilon(0)=0^+,
\qquad 0<\varepsilon\ll1.
\end{equation}
We assume that $\Theta$, $\dot\Theta$, $K_{ab}$, $a^a$ and their first derivatives
along $u^a$ remain finite as $\varepsilon\to0^+$, in accordance with the regular,
non-singular signature-change conditions (see e.g.\ Ref.~\cite{Hayward}).

\subsection{Geometric decomposition and small parameters}

We introduce two independent small, dimensionless parameters: $\sigma_K$ (anisotropy of
the extrinsic curvature) and $\sigma_a$ (non--geodesicity of $u^a$).

\smallskip
\paragraph*{Extrinsic curvature.}
The extrinsic curvature of the $u^a$--orthogonal slices is decomposed as
\begin{equation}
K_{mn}
  = \frac{K}{3}\,h_{mn} + \sigma_K\,\bar K_{mn},
\qquad
h^{mn}\,\bar K_{mn}=0,
\label{eq:K-decomp}
\end{equation}
where $K\equiv g^{mn}K_{mn}$ is the expansion scalar of the congruence, and
$\bar K_{mn}$ encodes the traceless anisotropic part. The limit
$\sigma_K\to 0$ corresponds to exact FRW isotropy at fixed $K$.

\smallskip
\paragraph*{Velocity field and acceleration.}
The preferred timelike congruence satisfies
\begin{equation}
g_{mn}u^m u^n = -1,
\qquad
a^m \equiv u^n\nabla_n u^m .
\end{equation}
We assume that $u^a$ is nearly geodesic,
\begin{equation}
a_m = \sigma_a\,\bar a_m,
\qquad
\bar a_m u^m=0,
\label{eq:a-decomp}
\end{equation}
with $\sigma_a\ll1$. In the limit $\sigma_a\to0$ the congruence becomes exactly
geodesic (comoving). The small parameters $(\sigma_K,\sigma_a)$ are independent of
$\varepsilon(t)$: they measure deviations from isotropy and geodesy, whereas
$\varepsilon$ measures the distance from the signature-change surface.

\subsection{Heat flux $q_m$}

From Sec.~\ref{EM tensor} the heat flux is
\begin{equation}
q_m = \Theta\,a^n K_{mn}.
\end{equation}
Substituting \eqref{eq:K-decomp} and \eqref{eq:a-decomp} yields
\begin{align}
q_m
 &= \Theta\,(\sigma_a \bar a^n)
    \left( \frac{K}{3} h_{mn} + \sigma_K \bar K_{mn} \right)
\nonumber\\[2pt]
 &= \sigma_a\,\Theta\,\frac{K}{3}\,\bar a_m
    + \sigma_a\sigma_K\,\Theta\,\bar a^n\bar K_{mn}.
\end{align}
Since $\Theta$, $K$, $\bar a_m$ and $\bar K_{mn}$ are assumed bounded near $\Sigma_0$, there exists a constant $C_q>0$ such that, in a local orthonormal frame,
\begin{equation}
|q_m| \le C_q\,\sigma_a
\qquad\Rightarrow\qquad
q_m = \mathcal{O}(\sigma_a),
\label{eq:q-scaling}
\end{equation}
Thus the heat flux is linearly suppressed by the near-geodesic parameter and vanishes
in the limit $\sigma_a\to0$.

\subsection{Anisotropic stress $\pi_{mn}$}

The anisotropic stress tensor derived in Sec.~\ref{EM tensor} is
\begin{align}
\pi_{mn} &= \Theta \Bigg[
      2 t_{(m} K_{n)a} \nabla_{\mathbf{u}} u^a
    - K K_{mn}
    - \nabla_{\mathbf{u}} K_{mn}
    \nonumber\\
&\qquad \qquad \qquad+ \frac{h_{mn}}{3}\bigl( \nabla_{\mathbf{u}} K + K^2 \bigr)
    \Bigg]
\nonumber\\
&\quad
    + \frac{\dot{\Theta}}{6}\left( K h_{mn} - 3 K_{mn} \right),
\label{eq:pi-def}
\end{align}
with $\nabla_{\mathbf u}\equiv u^a\nabla_a$. For an exactly isotropic case, it is straightforward to check that
$\pi_{mn}=0$ due to non-trivial cancellations among the terms in
\eqref{eq:pi-def}.

We now perturb around this isotropic, geodesic configuration. Define
\begin{equation}
\delta K_{ab} \equiv K_{ab} - \frac{K}{3}h_{ab} = \sigma_K\,\bar K_{ab},
\qquad
\delta a^c \equiv a^c = \sigma_a\,\bar a^c.
\end{equation}
Since \eqref{eq:pi-def} is polynomial in $K_{ab}$, $a^a$, $g_{ab}$, $\Theta$ and
$\dot\Theta$, we can expand $\pi_{mn}$ to first order in
$(\sigma_K,\sigma_a)$ around the isotropic, geodesic background:
\begin{equation}
\pi_{mn}
  = \sigma_K\,X_{mn} + \sigma_a\,Y_{mn}
    + \mathcal{O}(\sigma_K^2,\sigma_a^2,\sigma_K\sigma_a),
\label{eq:pi-linear}
\end{equation}
where $X_{mn}$ and $Y_{mn}$ are linear functionals of $\bar K_{ab}$ and
$\bar a^a$ whose coefficients are built from $K$, $\nabla_{\mathbf u}K$,
$\Theta$ and $\dot\Theta$. Crucially, there are no factors of $(1+\Theta)^{-1}$
in \eqref{eq:pi-def}, so under our regularity assumptions $X_{mn}$ and $Y_{mn}$
remain bounded as $\varepsilon\to0^+$, even though $\dot\Theta$ itself can be
numerically large. Thus there exists $C_\pi>0$ such that, in a local orthonormal
frame,
\begin{equation}
|\pi_{mn}| \le C_\pi\,(\sigma_K+\sigma_a)
\qquad\Rightarrow\qquad
\pi_{mn} = \mathcal{O}(\sigma_K+\sigma_a),
\label{eq:pi-scaling}
\end{equation}
uniformly for small $\varepsilon>0$.
The potentially large factor $\dot\Theta$ appears only in the bounded coefficients
contained in $C_\pi$; it does not introduce any $1/\varepsilon$ divergence.

\subsection{Comparison with the isotropic energy density}

From Sec.~\ref{EM tensor}, the effective energy density can be written as
\begin{equation}
\rho = \frac{F}{2}\left( \nabla_m a^m - {}^{(3)}\!R + \Theta\,\nabla_{\mathbf u}K\right),
\qquad
F \equiv \frac{\Theta}{1+\Theta}.
\label{eq:rho-appendix}
\end{equation}
Near $\Sigma_0$ we have $\Theta=-1+\varepsilon$ and
\begin{equation}
F = \frac{-1+\varepsilon}{\varepsilon}
  = -\frac{1}{\varepsilon} + 1 .
\end{equation}
Using $a^m=\sigma_a\bar a^m$ we obtain
\begin{equation}
\nabla_m a^m = \mathcal{O}(\sigma_a).
\end{equation}
It is convenient to introduce
\begin{equation}
\Delta \;\equiv\; \nabla_{\mathbf u}K + {}^{(3)}\!R,
\end{equation}
Hence
\begin{align}\label{eq:rho-bounds}
\rho
 &= \frac{1}{2}\Bigl(-\frac{1}{\varepsilon}+1\Bigr)
    \bigl(-\,\Delta + \nabla_m a^m + \varepsilon\,\nabla_{\mathbf u}K\bigr)
\nonumber\\[2pt]
 &= \frac{\Delta}{2\varepsilon}
    + \mathcal{O}\!\left(\frac{\sigma_a}{\varepsilon}\right)
    + \mathcal{O}(1),
\end{align}
where we have used the boundedness of $\nabla_{\mathbf u}K$ and
$\nabla_m a^m = \mathcal{O}(\sigma_a)$. 
Combining \eqref{eq:q-scaling}, \eqref{eq:pi-scaling} and \eqref{eq:rho-bounds},
and writing $\sigma\equiv\sigma_K+\sigma_a$ for brevity, we find
\begin{equation}
\frac{|q_a|}{\rho}
 \;\le\; \frac{2 C_q}{\Delta}\,\sigma_a\,\varepsilon
 \;=\; \mathcal{O}(\sigma\,\varepsilon),~~
\frac{|\pi_{ab}|}{\rho}
 \;\le\; \frac{2 C_\pi}{\Delta}\,\sigma\,\varepsilon
 \;=\; \mathcal{O}(\sigma\,\varepsilon),
\end{equation}
for $0<\varepsilon\ll1$.

These bounds make precise the statement quoted in Sec.~\ref{Inflation}:
the relative contribution of heat flux and anisotropic stress to the Einstein
equations is suppressed by $\sigma\,\varepsilon$ and therefore:
in this sense the leading dynamics near $\Sigma_0$ is governed by the
isotropic sector $(\rho,P)$, and the imperfect-fluid corrections from $q_a$ and
$\pi_{ab}$ are genuinely subleading.

\section{Other admissible interpolators }\label{app:interpolators}

\subsection{Tanh interpolator}

\paragraph{Definition and slope.}
\begin{equation}
\begin{aligned}
\Theta_{\tanh}(t;\lambda) &= \tanh(\lambda t)-1, \\
S(t) &= \dot\Theta(t)=\lambda\,\mathrm{sech}^{2}(\lambda t), \\
S(0^{+}) &= \lambda .
\end{aligned}
\end{equation}
The microscopic transition width scales as $\Delta t_{\rm tr}\sim \lambda^{-1}$.

\paragraph{Inflation window and end time.}
Inflation requires $S_{\rm th}<S(t)\le S_c$, i.e.
\begin{equation}
\lambda>\frac{S_c}{3}
\qquad \text{and} \qquad
\lambda\le S_c
\quad \text{(to avoid } w<-1).
\end{equation}
The end time is fixed by $S(t_{\rm end})=S_{\rm th}$:
\begin{equation}
\begin{aligned}
\mathrm{sech}^{2}\!\bigl(\lambda t_{\rm end}\bigr)
    &= \frac{S_{\rm th}}{\lambda}, \\[4pt]
\Delta t_{\rm inf}^{(\tanh)}
    &= \frac{1}{\lambda}\,
       \operatorname{arcosh}\!\left(
          \sqrt{\frac{\lambda}{S_{\rm th}}}
       \right)
     = \frac{1}{\lambda}\,
       \operatorname{arsech}\!\left(
          \sqrt{\frac{S_{\rm th}}{\lambda}}
       \right).
\end{aligned}
\label{eq:Dt-tanh-app}
\end{equation}

\subsection{Generalized logistic interpolator}

\paragraph{Definition and slope.}
To interpolate from Euclidean to Lorentzian across $\Theta=-1$ one may consider

\begin{equation}
\begin{aligned}
\Theta_{\rm GL}(t;a,b,m)
    &= \frac{2}{1+\exp\!\bigl[-a (t/b)^{m}\bigr]} - 2, \\[6pt]
&\quad a,b>0,\qquad
m\ \text{odd integer},
\end{aligned}
\label{eq:GL-def-odd}
\end{equation}
with derivative
\begin{equation}
\begin{aligned}
S(t) &\equiv \dot\Theta(t)
      = \frac{2am}{b^{m}}
        \frac{t^{\,m-1} e^{-a(t/b)^{m}}}
             {\bigl[1+e^{-a(t/b)^{m}}\bigr]^{2}}, \\[8pt]
S(0^{+}) &=
\begin{cases}
\dfrac{a}{2b}, & m=1,\\[6pt]
0, & m\ge 3\ \text{odd}.
\end{cases}
\end{aligned}
\label{eq:GL-S-odd}
\end{equation}
For $m=1$ one has
\begin{equation}
S(t)=\frac{a}{2b}\,\mathrm{sech}^2\!\left(\frac{a t}{2b}\right),
\end{equation}
which coincides with the tanh slope upon identifying $\lambda = a/(2b)$ in \eqref{eq:Dt-tanh-app}. Thus the $m=1$ generalized logistic interpolator is equivalent, at the level of $S(t)$ and $\Delta t_{\rm inf}$, to the tanh case. For odd $m\ge3$ the onset is delayed to a finite time $t>0$ and the inflationary duration $\Delta t_{\rm inf}$ can be obtained numerically from $S(t_{\rm end})=S_{\rm th}$.

\subsection{Power--law (arctan) interpolator}

\paragraph{Definition and slope.}
\begin{equation}
\begin{aligned}
\Theta_{\rm PL}(t;\lambda)
    &= \frac{2}{\pi}\arctan(\lambda t)-1, \\[4pt]
S(t)
    &= \dot\Theta(t)
     = \frac{2}{\pi}\,\frac{\lambda}{1+\lambda^{2}t^{2}}, \\[4pt]
S(0^{+})
    &= \frac{2}{\pi}\lambda .
\end{aligned}
\end{equation}
This profile has $\Theta(0)=-1$ and $\Theta\to 0^{-}$ as $t\to+\infty$.

\paragraph{Inflation window and end time.}
Inflation occurs only while
\begin{equation}
\frac{S_c}{3}<\frac{2}{\pi}\lambda\le S_c
\quad\Longleftrightarrow\quad
\frac{\pi}{2}\,\frac{S_c}{3}<\lambda\le \frac{\pi}{2}\,S_c.
\end{equation}
Solving $S(t_{\rm end})=S_{\rm th}$ gives
\begin{equation}
\begin{aligned}
\Delta t_{\rm inf}^{(\mathrm{PL})}
    &= \frac{1}{\lambda}
       \sqrt{\frac{S_{0}}{S_{\rm th}} - 1}, \\[4pt]
S_{0} &= S(0^{+}) = \frac{2}{\pi}\lambda, \\[4pt]
\Delta t_{\rm inf}^{(\mathrm{PL})}
    &= \frac{1}{\lambda}
       \sqrt{\frac{6\,\lambda}{\pi S_{c}} - 1}.
\end{aligned}
\label{eq:PL-duration}
\end{equation}
\begin{table}[h]
\centering
\renewcommand{\arraystretch}{1.2}
\resizebox{\columnwidth}{!}{%
\begin{tabular}{lccc}
\hline\hline
Interpolator & $\Theta(t)$ & $S(0^{+})$ & $\Delta t_{\rm inf}$ (closed form) \\
\hline
tanh &
$\tanh(\lambda t)-1$ &
$\lambda$ &
$\displaystyle \frac{1}{\lambda}\,
\operatorname{arcosh}\!\sqrt{\frac{\lambda}{S_{\rm th}}}$ \\[6pt]
generalized logistic ($m=1$) &
$\dfrac{2}{1+e^{-a t/b}}-2$ &
$\dfrac{a}{2b}$ &
$\displaystyle \frac{2b}{a}\,
\operatorname{arcosh}\!\sqrt{\frac{a}{2b\,S_{\rm th}}}$ \\[8pt]
power--law (arctan) &
$\dfrac{2}{\pi}\arctan(\lambda t)-1$ &
$\dfrac{2}{\pi}\lambda$ &
$\displaystyle \frac{1}{\lambda}\,
\sqrt{\frac{S_0}{S_{\rm th}}-1}$ \\
\hline\hline
\end{tabular}%
}
\caption{Smooth interpolators that realize an inflationary interval when 
$S_{\rm th}<S(0^{+})\le S_c$, with $S_{\rm th}\equiv S_c/3$. 
The expressions give the duration $\Delta t_{\rm inf}$ obtained from 
$S(t_{\rm end})=S_{\rm th}$, where $S_0\equiv S(0^+)$. For generalized logistic profiles with odd $m\ge3$, $S(0^{+})=0$ and the onset 
is delayed to a finite time $t>0$, so no simple closed-form expression for 
$\Delta t_{\rm inf}$ is available.}
\label{tab:interpolators-clean}
\end{table}
\section{Phenomenological e-fold estimates}\label{app:efolds}

Given an interpolator $\Theta(t)$ and a background geometry, the exact number of e-folds
accumulated during the geometric inflationary window is
\begin{equation}
N_e \equiv \ln\frac{a(t_{\rm end})}{a(t_\star)}
      = \int_{t_\star}^{t_{\rm end}} H(t)\,dt,
\label{eq:Ne-app-def}
\end{equation}
with $t_\star$ chosen just after the transition and $t_{\rm end}$ fixed by the geometric
condition $S(t_{\rm end})=S_{\rm th}$. This relation is purely kinematic and does not assume
any particular matter content or equation of state.

\subsection{Geometric scale and quasi--de Sitter regime}

In a closed FRW background with expansion rate $H$ and curvature ${}^{(3)}\!R=6/a^2$ one has
$K=3H$ and
\begin{equation}
\Delta = \nabla_u K + {}^{(3)}\!R = 3\dot H + \frac{6}{a^2}.
\end{equation}
It is convenient to introduce the geometric deviation parameter
\begin{equation}
\epsilon_G \equiv \frac{\Delta}{3H^2}
                = \frac{\dot H}{H^2} + \frac{2}{a^2 H^2},
\qquad
\Delta = 3\epsilon_G H^2,
\end{equation}
so that the geometric scale $S_c$ of Eq.~\eqref{eq:ScDef} reads
\begin{equation}
S_c = \frac{3\Delta}{2\varepsilon K}
    = \frac{3\,\epsilon_G\,H}{2\varepsilon}.
\end{equation}
Initially just after the transition the curvature term $2/(a^2 H^2)$ dominates and
$\epsilon_G>0$, ensuring $\Delta>0$ and hence positive effective energy density
$\rho\simeq\Delta/(2\varepsilon)$.

On the short geometric window we are interested in, we assume the background is quasi--de Sitter
in the sense that $H(t)$ varies slowly. Let $H_\star\equiv H(t_\star)$ and define
\begin{equation}
\epsilon_{\rm eff}\;\equiv\;
\sup_{t\in[t_\star,t_{\rm end}]}\left|\frac{\dot H(t)}{H_\star^2}\right|,
\end{equation}
so that by construction
\begin{equation}
|\dot H(t)|\;\le\;\epsilon_{\rm eff}\,H_\star^2,
\qquad t\in[t_\star,t_{\rm end}].
\end{equation}
Writing
\begin{equation}
H(t) = H_\star + \delta H(t),
\end{equation}
we can split Eq.~\eqref{eq:Ne-app-def} as
\begin{equation}
N_e = H_\star \Delta t_{\rm inf}
    + \int_{t_\star}^{t_{\rm end}}\delta H(t)\,dt,
\qquad
\Delta t_{\rm inf}\equiv t_{\rm end}-t_\star.
\label{eq:Ne-split-app}
\end{equation}

Integrating the bound on $\dot H$ once gives
\begin{equation}
|\delta H(t)| \;=\; \left|\int_{t_\star}^t \dot H(\tau)\,d\tau\right|
   \;\le\; \epsilon_{\rm eff}\,H_\star^2|t-t_\star|,
\end{equation}
and hence
\begin{align}
\left| \int_{t_\star}^{t_{\rm end}} \delta H(t)\, dt \right|
&\le \int_{t_\star}^{t_{\rm end}} |\delta H(t)|\, dt
 \;\le\; \epsilon_{\rm eff}\, H_\star^2
        \int_{t_\star}^{t_{\rm end}} |t - t_\star|\, dt
\nonumber\\[4pt]
&= \frac{1}{2}\, \epsilon_{\rm eff}\, H_\star^2\, \Delta t_{\rm inf}^2 .
\end{align}

Therefore
\begin{equation}
N_e = H_\star \Delta t_{\rm inf}
    + \mathcal O\!\left(\epsilon_{\rm eff}\,H_\star^2 \Delta t_{\rm inf}^2\right),
\label{eq:Ne-approx-app}
\end{equation}
so that, using $N_e\simeq H_\star \Delta t_{\rm inf}$ in the leading term, the \emph{relative}
correction is parametrically of order
\begin{equation}
\frac{\delta N_e}{N_e}
\;\sim\;\epsilon_{\rm eff}\,H_\star \Delta t_{\rm inf}
\;\sim\;\epsilon_{\rm eff}\,N_e.
\end{equation}
Provided $\epsilon_{\rm eff}N_e\ll1$ (for instance $\epsilon_{\rm eff}\lesssim 10^{-3}$ when
$N_e\sim 50$--$60$), the approximation $N_e\simeq H_\star\Delta t_{\rm inf}$ is accurate at the
few-percent level. Importantly, Eq.~\eqref{eq:Ne-approx-app} does not assume a constant equation
of state; it follows only from the short duration and quasi--de Sitter character of the geometric
window.

\subsection{Constant-\texorpdfstring{$\epsilon_H$}{epsilonH} toy model}

As a simple check of the quasi--de Sitter estimate, consider a purely kinematic toy model with constant
\begin{equation}
\epsilon_H \equiv -\frac{\dot H}{H^2} = \text{const},
\end{equation}
so that 
\(
\dot H = -\epsilon_H H^2
\)
and hence
\(
H(t) = H_\star/[1+\epsilon_H H_\star (t-t_\star)].
\)
Substituting into Eq.~\eqref{eq:Ne-app-def} gives
\begin{equation}
N_e = \frac{1}{\epsilon_H}\,
\ln\!\bigl(1 + \epsilon_H H_\star \Delta t_{\rm inf}\bigr),
\qquad
\Delta t_{\rm inf} \equiv t_{\rm end}-t_\star.
\end{equation}
Introducing $x \equiv \epsilon_H H_\star \Delta t_{\rm inf}$, one has
\begin{equation}
N_e = \frac{1}{\epsilon_H}\,\ln(1+x),
\qquad
N_e^{\rm (approx)} \equiv H_\star \Delta t_{\rm inf} = \frac{x}{\epsilon_H},
\end{equation}
so that
\begin{equation}
\frac{N_e}{N_e^{\rm (approx)}} = \frac{\ln(1+x)}{x}
    = 1 - \frac{x}{2} + \mathcal O(x^2).
\end{equation}
For the expression values used in the main text ($x\simeq 0.06$--$0.08$ for the interpolators considered), this ratio differs from unity by only a few percent, in agreement with the general bound \eqref{eq:Ne-approx-app}.

\subsection{Summary of geometric scaling}

Collecting the frozen-threshold expression results, the two representative interpolators yield
\begin{equation}
N_e^{(\tanh)} \;\approx\; 0.76\,\frac{\varepsilon}{\epsilon_G},
\qquad
N_e^{(\mathrm{PL})} \;\approx\; 0.60\,\frac{\varepsilon}{\epsilon_G},
\end{equation}
up to absolute corrections of order
$\epsilon_{\rm eff} H_\star^2\Delta t_{\rm inf}^2\sim\epsilon_{\rm eff}N_e^2$, i.e.\ relative
corrections $\sim\epsilon_{\rm eff}N_e$. These formulas are controlled analytic limits in which the geometric threshold is treated as approximately fixed. They are useful for displaying the dependence on the interpolator and the local geometric scale, but they should not be interpreted as the general curvature-dominated e--fold prediction. When the curvature term in $\Delta$ is initially important, ${\cal K}(N)$ evolves according to Eq.~\eqref{eq:K-moving-threshold-app}, and the exact duration is set by the interval over which $S_{\rm th}(N)<S(N)\le S_c(N)$. This preserves the local geometric criterion while making clear that the large-e--fold question is a moving-threshold problem (see \ref{moving threshols}) rather than a statement about the initial curvature alone.

\subsection{Remark on moving thresholds} \label{moving threshols}

The closed-form estimates above treat $S_{\rm th}$ and $S_c$ as approximately fixed over the interval during which the interpolator slope is compared with the threshold. This is a useful local analytic expression, but in a curvature-supported closed branch the threshold can move because the curvature term in
\begin{equation}
\Delta=3\dot H+\frac{6}{a^2}
\end{equation}
evolves during expansion. To make this explicit, introduce
\begin{equation}
\epsilon_H\equiv-\frac{\dot H}{H^2},
\qquad
{\cal K}\equiv\frac{1}{a^2H^2}.
\end{equation}
Then
\begin{equation}
\epsilon_G\equiv\frac{\Delta}{3H^2}
=
-\epsilon_H+2{\cal K},
\end{equation}
and hence
\begin{equation}
S_{\rm th}(N)
=
\frac{H(N)}{2\varepsilon(N)}
\left[2{\cal K}(N)-\epsilon_H(N)\right],
\qquad
S_c(N)=3S_{\rm th}(N).
\label{eq:moving-threshold-app}
\end{equation}
The curvature fraction obeys the exact kinematic relation
\begin{equation}
\frac{d\ln{\cal K}}{dN}
=
-2(1-\epsilon_H),
\end{equation}
so that
\begin{equation}
{\cal K}(N)
=
{\cal K}_\star
\exp\!\left[-2N+2\int_0^N\epsilon_H(N')\,dN'\right].
\label{eq:K-moving-threshold-app}
\end{equation}
In the almost de--Sitter limit, $\epsilon_H\simeq0$, this reduces to ${\cal K}(N)\simeq{\cal K}_\star e^{-2N}$. Thus, if the initial curvature contribution is large, the threshold is generally not fixed over many e--folds.

For a branch with approximately constant $\epsilon_H=\beta$, one has
\begin{equation}
H(N)=H_\star e^{-\beta N},
\qquad
{\cal K}(N)={\cal K}_\star e^{-2(1-\beta)N}.
\end{equation}
For the tanh interpolator, define
\begin{equation}
\alpha\equiv\frac{\lambda}{H_\star},
\qquad
x(N)\equiv\lambda t(N)=
\begin{cases}
\dfrac{\alpha}{\beta}\left(e^{\beta N}-1\right), & \beta\ne0,\\[6pt]
\alpha N, & \beta=0.
\end{cases}
\end{equation}
Then
\begin{equation}
\varepsilon_{\tanh}(N)=\tanh x(N),
\qquad
\frac{S_{\tanh}(N)}{H(N)}=
\alpha e^{\beta N}{\rm sech}^2x(N),
\end{equation}
while
\begin{equation}
\frac{S_{\rm th}(N)}{H(N)}=
\frac{2{\cal K}_\star e^{-2(1-\beta)N}-\beta}
{2\tanh x(N)}.
\end{equation}
The moving non--phantom band is determined by
\begin{equation}
1<{\cal R}_{\tanh}(N)\le3,
\qquad
{\cal R}_{\tanh}(N)\equiv\frac{S_{\tanh}(N)}{S_{\rm th}(N)},
\end{equation}
with
\begin{equation}
{\cal R}_{\tanh}(N)=
\frac{
2\alpha e^{\beta N}\tanh x(N)\,{\rm sech}^2x(N)
}
{
2{\cal K}_\star e^{-2(1-\beta)N}-\beta
}.
\label{eq:R-tanh-moving-app}
\end{equation}
The e--fold count in this moving-threshold description is $N_e=N_+-N_-$, where $N_-$ and $N_+$ are the first and last points of the connected interval satisfying $1<{\cal R}_{\tanh}(N)\le3$. In the almost de--Sitter limit this reduces to
\begin{equation}
{\cal R}_{\tanh}(N)=
\frac{
\alpha e^{2N}\tanh(\alpha N)\,{\rm sech}^2(\alpha N)
}
{{\cal K}_\star}.
\end{equation}
This shows explicitly how curvature dilution shifts the threshold even when $\epsilon_H$ is held fixed. Consequently the estimates $N_e^{(\tanh)}\simeq0.76\,\varepsilon/\epsilon_G$ and $N_e^{({\rm PL})}\simeq0.60\,\varepsilon/\epsilon_G$ should be understood as frozen-threshold expressions. In the curvature-dominated regime the large-e--fold question is instead controlled by the persistence of the moving band $S_{\rm th}(N)<S(N)\le S_c(N)$.

\bibliography{apssamp}

\end{document}